\documentclass[journal]{IEEEtran} 

\usepackage{xcolor}
\usepackage{graphicx}
\usepackage{subcaption}
\usepackage{amsthm}
\usepackage{amsmath}
\usepackage{amssymb}
\usepackage{enumerate}
\usepackage{physics}
\usepackage{caption}
\usepackage{cite}
\usepackage{tabularx}
\newtheorem{remark}{Remark}
\newtheorem{lemma}[remark]{Lemma}
\newtheorem{theorem}[remark]{Theorem}
\newtheorem{algorithm}{Algorithm}
\newtheorem{assumption}{Assumption}
\newtheorem{method}{Method}

\def\hP{\hat{P}}
\def\bP{\bar{P}}
\def\tS{\widetilde{S}}
\def\tP{\widetilde{P}}
\def\blue{\textcolor{black}}
\def\PP{\mathbb{P}} 
\def\EE{\mathbb{E}} 
\def\Id{\mathrm{I}} 
\newcommand{\prob}[1]{\mathbb{P}\left\{#1\right\}}
\newcommand{\Ex}[2][]{\mathbb{E}_{#1}\left\{#2\right\}}
\allowdisplaybreaks[4]

\begin{document}
\title{Stochastic MPC with Dynamic Feedback Gain Selection and Discounted Probabilistic Constraints}
\author{Shuhao~Yan, Paul J. Goulart and~Mark~Cannon
\thanks{The authors are with the Department of Engineering Science, University
of Oxford, OX1 3PJ, UK. (E-mail: shuhao.yan@eng.ox.ac.uk; paul.goulart@eng.ox.ac.uk; mark.cannon@eng.ox.ac.uk)}}
\markboth{Manuscript}{?}

\maketitle
\begin{abstract}
This paper considers linear discrete-time systems with additive disturbances, and designs a Model Predictive Control (MPC) law incorporating a dynamic feedback gain to minimise a quadratic cost function subject to a single chance constraint. The feedback gain is selected online 
and we provide two selection methods based on minimising upper bounds on predicted costs. The chance constraint is defined as a discounted sum of violation probabilities on an infinite horizon. By penalising violation probabilities close to the initial time and assigning violation probabilities in the far future with vanishingly small weights, this form of constraints allows for an MPC law with guarantees of recursive feasibility
without a boundedness assumption on the disturbance. A computationally convenient MPC optimisation problem is formulated using Chebyshev's inequality and we introduce an online constraint-tightening technique to ensure recursive feasibility.
The closed loop system is guaranteed to satisfy the chance constraint and a quadratic stability condition. With dynamic feedback gain selection, the closed loop cost is reduced and conservativeness of Chebyshev's inequality is mitigated. Also, a larger feasible set of initial conditions can be obtained. Numerical simulations are given to show these results.
\end{abstract}

\begin{IEEEkeywords}
Model predictive control, chance constraints, Chebyshev inequality, dynamic programming, multiobjective optimisation, stochastic convergence.
\end{IEEEkeywords}
\IEEEpeerreviewmaketitle

\section{Introduction} \label{section:introduction}
\IEEEPARstart{R}{obust} control methods for systems with unknown disturbances must take into account worst-case disturbance bounds in order to guarantee satisfaction of hard constraints on system states and control inputs~\cite{Mayne2000auto}. However, for problems with stochastic disturbances and constraints that are allowed to be violated up to a specified probability, worst-case control strategies can be unnecessarily conservative. This motivated the development of stochastic MPC, which addresses optimal control problems for systems with chance constraints by making use of information on the distribution of model uncertainty~\cite{KOUVARITAKIS2010auto}.

Available methods for approximating chance constraints include analytical approximation and sampling methods. The former aims to provide tractable deterministic optimisation problems while the latter generally results in randomised methods. In \cite{Farina13}, Cantelli's inequality is used to turn the chance constraint on states into linear constraints. In~\cite{Hashimoto13} and \cite{Schildbach2015auto}, Chebyshev's inequality is used to reformulate chance constraints as a quadratic and a linear matrix inequality constraint, respectively. These two inequalities can handle a wide range of probability distributions, and only require information on the first and second moments of additive disturbance distributions. However, the resulting approximate chance constraints only provide tight bounds for specific probability distributions and otherwise are  conservative.
A scenario approach is used in~\cite{schildbach2014scenario} to impose time-average expectation constraints on system states. Although constraint satisfaction is demonstrated for the closed loop system, recursive feasibility of online optimisations are assumed but not ensured. More generally, sample based methods are unable to ensure recurrence of feasibility of receding horizon optimisation problems unless they are combined with robust bounds on model uncertainty. For example, a recursively feasible MPC strategy is proposed in \cite{j_77}, with chance constraints imposed using a scenario approach at the first prediction time step and replaced by robust constraints at later times.

In order to provide guarantees of recursive feasibility and constraint satisfaction in closed loop operation with reduced conservativeness, online constraint tightening techniques are proposed in \cite{korda14,Lorenzen17ieeetr,fleming17}. These methods rely on knowledge of worst-case disturbance bounds, and their degree of conservativeness increases as the disturbance bounds become more conservative.
An adaptive approach is developed in \cite{munoz2018stochastic} which aims to avoid this problem by introducing a scaling factor for tightening parameters that is adapted online, making use of past observations of constraint violations. The authors show that the time-average constraint violation rate converges in probability to a specified limit. For the case that bounds are known on disturbances, hard constraints on control inputs can be incorporated. However, without bounded disturbance distributions or an assumption that the open loop system is stable (as in \cite{hokayem2009stochastic}, for example), it is not possible to guarantee satisfaction of hard constraints. \blue{To ensure recursive feasibility under unbounded additive disturbances, previous work (e.g.\ \cite{Farina13,hewing2018stochastic,hewing2020recursively}) has resorted to a backup initialisation of the MPC optimisation problem when infeasibility occurs. This
design can prevent feedback from the actual state measurement and could thus yield inadequate performance.}

For problems involving stochastic uncertainty, the optimal expected value of predicted cost is typically used to perform a Lyapunov analysis of closed loop stability.
A performance metric widely used in this setting is the long-run expected average cost \cite{Chatterjee15Sstability}. Although the vast majority of stability results are derived by imposing terminal constraints, there are a number of alternative approaches (e.g. \cite{lorenzen2019stochastic,limon2006stability,maciejowski2000robust}).

Discounted costs and constraints are present in many stochastic control settings (e.g. \cite{blackwell1965discounted,Bertsekas:2000,VanParys_discounted_cost_bounds,kamgarpour17}), as well as in reinforcement learning \cite{bertsekas2019reinforcement}, financial engineering (e.g.  \cite{barratt2020multi,nystrup2019multi,boyd2014performance}) and ecosystem management (e.g. \cite{clark1973profit,ludwig2005uncertainty}).
Discount factors in optimal control problems allow performance in the near future to be prioritised over long-term behaviour. This shift of emphasis is vital for ensuring recursive feasibility of chance-constrained control problems involving possibly unbounded disturbances.
In Dynamic Programming (e.g.~\cite{blackwell1965discounted,Bertsekas:2000}), discounting is commonly employed to ensure that infinite horizon problems with possibly unbounded cost per stage are well-defined.
In economics, discounting allows aggregation of current and potential future costs and revenues. For example, \cite{clark1973profit} shows that varying discount factors on future revenue can affect harvesting policies.


In contrast to existing work on stochastic MPC,
this paper considers
linear discrete-time systems subject to possibly unbounded additive disturbances. We propose an MPC strategy incorporating a dynamic feedback gain to minimise a quadratic cost while satisfying a chance constraint. The constraint combines long-term and short-term considerations by imposing a bound on discounted violation probabilities accumulated over an infinite horizon. The chance constraint is reformulated using Chebyshev's inequality using knowledge of only the first and second moments of the disturbance input to obtain a convex optimisation problem. The main features and contributions of this paper are summarised as follows:
\begin{itemize}
\item We use a discount factor to ensure that the chance constraint is well-defined and to prioritise near-future system behaviour over steady state performance.
\item A constraint tightening technique is proposed to ensure recursive feasibility of online MPC optimisation problems and constraint satisfaction in closed loop without requiring disturbances to be bounded.
\item The closed loop system satisfies a quadratic stability condition without the need for terminal constraints.
\end{itemize}

This paper extends preliminary results that appeared in~\cite{Yan_discountsmpc_18} by reducing the conservativeness of Chebyshev's inequality using a dynamic feedback gain in the definition of the MPC law. 
This is achieved by solving offline a set of multiobjective optimisation problems with the cost and constraint functions considered as conflicting objectives to generate a set of feedback gains providing various trade-offs between these two objectives.
Two methods of online gain selection are proposed, minimising upper bounds on the optimal predicted cost while retaining guarantees of recursive feasibility and computationally simple MPC optimisation.
We show that the gain selection procedure can be configured so that the feedback gain converges almost surely to the unconstrained LQ-optimal solution, and the set of admissible initial conditions can be enlarged by choosing an appropriate initial gain. The MPC algorithm significantly improves closed loop performance in terms of the long-run expected average cost and reduces the conservativeness of closed loop constraint handling compared to MPC laws based on fixed feedback gains.

The paper is organised as follows.
The control problem is described and the controller structure is formulated in Section~\ref{section: problem description}.
Section~\ref{section:recursive feasibility} proposes an online constraint-tightening method for guaranteeing recursive feasibility.
Section~\ref{section:multiobjective optimisation and dynamic programming}
addresses multiobjective optimisation problems, solutions of which provide strictly stabilising feedback gains.
These feedback gains are used to define predicted control sequences via the dynamic feedback gain selection methods proposed in Section~\ref{section:dynamic selection}.
Section~\ref{section:smpc algorithm} summarises the proposed MPC algorithm and derives a bound on closed loop performance.
In Section~\ref{section:constraint satisfaction}, the closed loop behaviour of the tightening parameters is analysed and constraint satisfaction is proved.
Section~\ref{section:example} gives a numerical example illustrating the results obtained and the paper is concluded in Section~\ref{section:conclusion}.
Some proofs are given in the Appendix to improve readability.

\textit{Notation}:
The Euclidean norm of a vector $x\in\mathbb{R}^n$ is denoted $\| x\|$ and we define $\norm{x}^2_{Q} := x^\top Qx$.
The notation $Q\succcurlyeq0$ and $R\succ0$ indicates that $Q$ and $R$ are symmetric positive semidefinite and symmetric positive definite matrices respectively, and $\tr(Q)$ denotes the trace of $Q$.
A matrix of suitable dimension with all entries being $0$ is denoted $\mathbf{0}$, and an identity matrix of suitable dimension is denoted $\Id$.
The conditional probability of an event $\mathcal{F}$ given the state $x_k$ is denoted $\PP\{\mathcal{F} | x_k\} = \PP_k \{ \mathcal{F} \} $, the conditional expectation of $y$ given $x_k$ is $\EE \{y | x_k \} = \EE_k\{y\}$, and $\PP\{\mathcal{F}\}$, $\EE\{y\}$ are equivalent to $\PP_0 \{ \mathcal{F}\}$, $\EE_0 \{y\}$ respectively.
The sequence $\{ x_0,\ldots,x_{N-1}\}$ is denoted $\{x_i\}_{i=0}^{N-1}$. We denote the value of a variable $x$ at time $k$ as $x_k$, and the $i$-step-ahead predicted value of $x$ at time $k$ is denoted $x_{i|k}$.

\section{Problem Description} \label{section: problem description}
Consider an uncertain linear system
\begin{align}
x_{k+1}=Ax_{k}+Bu_{k}+\omega_k, \label{eqn:system dynamics:original state space model}
\end{align}
where $x_{k}\in \mathbb{R}^{n_x}$, $u_{k}\in \mathbb{R}^{n_u}$ are the system state and control input respectively. The unknown disturbance input $\omega_k \in \mathbb{R}^{n_x}$ is independently and identically distributed (i.i.d.) with known first and second moments
\begin{equation}
  \Ex{\omega_k}=\mathbf{0}, \quad \Ex{\omega_k\omega^\top_{k}}=:\Omega \succ 0.  \label{eq:disturbance statistics}
\end{equation}
The disturbance distribution may have infinite support, unlike the approach of \cite{Lorenzen17ieeetr}, which assumes the additive disturbance lies in a compact set.
We assume that the system state is measured directly and is available to the controller at each sample instant.

The system \eqref{eqn:system dynamics:original state space model} is subject to the constraint
\begin{equation}
\sum_{k=0}^\infty \gamma^k \mathbb{P}\left\{ \norm{C x_k + D u_k} \geq 1 \right\}  \leq e
\nonumber
\end{equation}
for given $C \in \mathbb{R}^{n_c\times n_x}$, $D \in \mathbb{R}^{n_c\times n_u}$, a positive scalar $e$ and a discount factor $\gamma \in (0,1)$.
To simplify presentation we set $D=0$ for the remainder of the paper, noting that all of the results given in Sections~III-VII apply to the case of non-zero $D$, and in the sequel the constraint
\begin{equation}
\sum_{k=0}^\infty \gamma^k \mathbb{P}\left\{ \norm{C x_k} \geq 1 \right\}  \leq e \label{eqn:constraint:probability constraint of the original form}
\end{equation}
is considered. We refer to $\mathbb{P}\{\| C x_k \|\geq 1\}$ as a \textit{violation probability}.

In this work, we design a controller to solve the following problem
\begin{equation}
\begin{aligned}
\min~&\EE \Bigl\{\sum_{k=0}^\infty \norm{x_k}^2_Q+ \norm{u_k}^2_R\Bigr\} \\
&\text{s.t. } \eqref{eqn:constraint:probability constraint of the original form} .
\end{aligned} \label{eqn:cost:original form of the cost function}
\end{equation}
The weighting matrices in the cost function of problem~\eqref{eqn:cost:original form of the cost function} are assumed to satisfy $Q\succcurlyeq0$ and $R\succ0$.

\begin{assumption}\label{assumption: controllability and observability}
$(A,B)$ is controllable and $(A,Q^{\frac{1}{2}})$ is observable.
\end{assumption}

The discounting in this problem introduces a special feature
that the probabilities of violating the condition $\| C x_k  \| < 1$ at time instants $k$ nearer to the initial time are weighted more heavily than those in the far future.
Therefore the discount factor allows for a trade-off between short-term and long-term behaviours and essentially determines how much priority the algorithm gives to constraint violations in the immediate future relative to those in the distant future.

\subsection{Finite horizon formulation and constraint handling}\label{sec:finite horizon formulation}
Problem \eqref{eqn:cost:original form of the cost function} employs an infinite horizon in the definition of both the cost and constraint.
This optimisation problem is computationally intractable since it involves an infinite sequence of decision variables, namely the control inputs $\{u_k\}_{k=0}^\infty$.
However, the use of an infinite horizon can impart desirable properties, notably stability \cite{Mayne2000auto}. It is therefore beneficial to formulate a similar problem with a cost function and a constraint that are defined over a finite horizon in such a way that they appropriately approximate the infinite horizon cost and constraint of the original problem. Moreover,
the probability distribution of states may be unknown at each time step and the evaluation of \textit{violation probabilities} is therefore generally intractable.
Even if the probability distribution of $\omega_k$ is known explicitly, computing a finite horizon version of \eqref{eqn:constraint:probability constraint of the original form}
 requires the solution of a set of multivariate convolution integrals, which is difficult to manage in general \cite{KOUVARITAKIS2010auto}.

We therefore propose to solve problem \eqref{eqn:cost:original form of the cost function} using a receding horizon approach wherein our control law is parameterised at each stage with a finite number of decision variables, and the constraint \eqref{eqn:constraint:probability constraint of the original form} is approximated conservatively using the two-sided Chebyshev inequality.  The resulting optimisation problem to be solved at each time step is then both finite dimensional and computationally tractable.

\subsection{Predicted nominal control input and state sequences} \label{sec:predicted sequence}
Before deriving the finite horizon expressions of the cost and the constraint as mentioned in the previous section, this section defines predicted nominal control input and state sequences.

The sequence of nominal control inputs predicted at time $k$ is given by
\begin{align}
&\bar{u}_{i|k}=K_k \bar{x}_{i|k} + c_{i|k},\quad i=0,\ldots,N-1 \label{eqn:nominal control input:first mode} \\
&\bar{u}_{N+i|k}=K_k \bar{x}_{N+i|k}, \quad i=0,1,\ldots, \label{eqn:nominal control input:second mode}
\end{align}
where $\bar{x}_{i|k}$ is the $i$-step-ahead prediction of the nominal state given information at time $k$, that is, $\Ex[k]{x_{i|k}}=\bar{x}_{i|k}$. The matrix-valued term $K_k$
is a stabilising feedback gain that is selected online at time $k$ from amongst a precomputed set of candidates.  The offline generation of candidates and the procedure used to select from them online will be detailed in Sections~\ref{section:multiobjective optimisation and dynamic programming} and \ref{section:dynamic selection}, respectively.
After selecting a gain $K_k$, the \textit{perturbation sequence}
$\{c_{0|k}, c_{1|k}, \ldots, c_{N-1|k}\}$ then constitutes the decision variables in the MPC optimisation problem to be solved at time $k$.


Given the predicted nominal control law \eqref{eqn:nominal control input:first mode}-\eqref{eqn:nominal control input:second mode}, the predicted nominal state trajectory is given by $\bar{x}_{0|k} = x_k$ and
\begin{align}
& \bar{x}_{i|k}= \Phi_k^i\bar{x}_{0|k}+\sum_{j=0}^{i-1} \Phi_k^{i-1-j}Bc_{j|k} , & i&=1,\ldots,N,
                                                                              \label{eqn:nominal state:dynamics describing the evolution of nominal state}\\
& \bar{x}_{N+i|k}= \Phi_k^i\bar{x}_{N|k}, &
                                                                i&=1,2,\ldots,
                                                                   \label{eqn:nominal state:dynamics describing the evolution of nominal state after time N}
\end{align} where $\Phi_k:=A+BK_k$.
The covariance matrix, $X_{i|k}$, of the $i$-step-ahead predicted state
is given by $X_{0|k} = \mathbf{0}$ and
\begin{equation}
X_{i|k}=\sum_{j=0}^{i-1} \Phi_k^j \Omega\left(\Phi_k^j\right)^\top ,\quad i=1,2,\ldots. \label{eqn:covariance matrix of state:expression of covariance matrix in term of sums involving W}
\end{equation}
We rewrite \eqref{eqn:nominal state:dynamics describing the evolution of nominal state} in a compact form as
\begin{align}
\mathbf{\bar{x}}_k&\!=\!M_x\left(K_k\right) \bar{x}_{0|k}+M_c\left(K_k\right) \mathbf{c}_k, \label{eqn:compact form of nominal system dynamics}\\
\begin{bmatrix}
\bar{x}_{1|k}\\
\bar{x}_{2|k}\\
\vdots\\
\bar{x}_{N|k}
\end{bmatrix}
&\!=\!
\begin{bmatrix}
\Phi_k\\
\Phi_k^2\\
\vdots\\
\Phi_k^N
\end{bmatrix}
\!\bar{x}_{0|k}
\!+\!
\begin{bmatrix}
B&&\\
\Phi_k B&&\\
\vdots&\ddots&\\
\Phi_k^{N-1}B&\cdots&B
\end{bmatrix}\!
\begin{bmatrix}
c_{0|k}\\
c_{1|k}\\
\vdots\\
c_{N-1|k}
\end{bmatrix}\!, \label{eqn:matrix form of nominal system dynamics}
\end{align}
and $M_x\left(K_k\right) \in \mathbb{R}^{Nn_x \times n_x}$, $M_c\left(K_k\right) \in \mathbb{R}^{Nn_x \times Nn_u}$. For simplicity we write these two matrices as $M_x$ and $M_c$, with the understanding that they depend on $K_k$.
\subsection{Online MPC optimisation}
Based on predicted sequences defined in Section~\ref{sec:predicted sequence} and employing Chebyshev's inequality, we give finite horizon expressions of the cost and the constraint and formulate an MPC optimisation problem to be solved repeatedly online.

Minimising the predicted cost
$\EE_k\{ \sum_{i=0}^\infty \norm{x_{i|k}}^2_Q\!+\! \norm{u_{i|k}}^2_R \}$
at time $k$ over the optimisation variable $\mathbf{c}_k$ is equivalent to minimising the cost defined in terms of the predicted nominal input sequence \eqref{eqn:nominal control input:first mode}-\eqref{eqn:nominal control input:second mode} and state trajectory \eqref{eqn:nominal state:dynamics describing the evolution of nominal state}-\eqref{eqn:nominal state:dynamics describing the evolution of nominal state after time N} by
\begin{align}
J(\bar{x}_{0|k},K_k,\mathbf{c}_k) \!:=\!\!\sum_{i=0}^{N-1}\!\Bigl(\norm{\bar{x}_{i|k}}^2_Q + \norm{K_k \bar{x}_{i|k}\!+\!c_{i|k}}^2_R\Bigr) \nonumber \\
+\norm{\bar{x}_{N|k}}^2_{P_k}.
\label{eqn:cost function: explicit cost function, final form}
\end{align}
Here $\norm{ \bar{x}_{N|k} }^2_{P_k}$
is the terminal cost and $P_k \succ 0$ is chosen as the solution of
\begin{equation}
P_k = Q+K_k^\top RK_k + \Phi_k^\top  P_k\Phi_k. \label{eqn:lyap P_k}
\end{equation}
Using \eqref{eqn:compact form of nominal system dynamics}-\eqref{eqn:matrix form of nominal system dynamics}, we rewrite
\eqref{eqn:cost function: explicit cost function, final form} in a compact form  as
\begin{equation}
J(\bar{x}_{0|k},K_k,\mathbf{c}_k)=
\begin{bmatrix}
   \bar{x}_{0|k}\\
   \mathbf{c}_k
   \end{bmatrix}^\top  \!\!
   W_2\left(K_k\right)
   \begin{bmatrix}
   \bar{x}_{0|k}\\
   \mathbf{c}_k
   \end{bmatrix}, \label{eqn:def W2}
\end{equation}
where $W_2\left(K_k\right)$ is a function of $K_k$ and its expression is omitted here for simplicity.

Approximating the LHS of \eqref{eqn:constraint:probability constraint of the original form} at time $k$ \blue{by its upper bound derived from} a direct application of the two-sided Chebyshev inequality\blue{\cite[Section 7.3]{grimmett2001probability}}, we obtain
\begin{equation}
\sum^{\infty}_{i=0}\gamma^i \bigl( \norm{ C\bar{x}_{i|k}}^2 + \tr (   C^\top C X_{i|k} ) \bigr)
\leq \varepsilon_k. \label{eqn:constraint: deterministic online constraint by chebyshev inequality}
\end{equation}
Here $\varepsilon_k$ replaces $e$ in \eqref{eqn:constraint:probability constraint of the original form} as a threshold on the resulting constraint function, and it is a design parameter to be chosen at time $k$ in some way such that we can ensure recursive feasibility of online MPC optimisations under possibly unbounded disturbances.
The advantages of this approach are that it can cope with unknown disturbance probability distributions with known first and second moments, and furthermore it results in a quadratic constraint that is straightforward to implement.
By the following lemma, we show that the LHS of \eqref{eqn:constraint: deterministic online constraint by chebyshev inequality} is equivalent to a finite horizon expression.
\begin{lemma} \label{lemma: expression for terminal term in constraint}
  Let $\tP_k$ be the solution of
  \begin{equation}
\tP_k = \gamma\Phi_k^\top  \tP_k \Phi_k + C^\top C \label{eq:Plyap}.
  \end{equation}
Then
\begin{align*}
\sum_{i=N}^{\infty} \gamma^i \bigl\| C\bar{x}_{i|k}\bigr\|^2  &=\gamma^N \bigl\| \bar{x}_{N|k} \bigr\|^2_{\tP_k}, \\
\sum^{\infty}_{i=0} \gamma^i \tr \bigl(  C^\top C X_{i|k} \bigr) &=  \frac{\gamma}{1-\gamma} \tr\bigl( \Omega\tP_k \bigr),
\end{align*}
where $\bar{x}_{i|k}$ is given by~\eqref{eqn:nominal state:dynamics describing the evolution of nominal state after time N} for all $i\geq N$, $X_{0|k}=\bf{0}$ and $X_{i|k}$ is given by \eqref{eqn:covariance matrix of state:expression of covariance matrix in term of sums involving W} for all $i\geq 1$.
\end{lemma}

By Lemma~\ref{lemma: expression for terminal term in constraint},
\eqref{eqn:constraint: deterministic online constraint by chebyshev inequality} is equivalent to
\begin{equation}
\sum^{N-1}_{i=0}\gamma^i \norm{ C\bar{x}_{i|k}}^2 \!+ \gamma^N \norm{\bar{x}_{N|k}}^2_{\tP_k} \!+\!  \frac{\gamma}{1-\gamma}\tr\bigl( \Omega\tP_k \bigr) \leq \varepsilon_k , \label{eqn:constraint: short expression of deterministic online constraint}
\end{equation}
where $\gamma^N \norm{\bar{x}_{N|k}}^2_{\tP_k}$ is the terminal term of the infinite discounted sum
associated with predicted nominal states and $\frac{\gamma}{1-\gamma}\tr\bigl( \Omega\tP_k \bigr)$ is the infinite discounted sum associated with covariance, which remains finite due to the discount factor.
Using \eqref{eqn:compact form of nominal system dynamics}-\eqref{eqn:matrix form of nominal system dynamics}, we rewrite \eqref{eqn:constraint: short expression of deterministic online constraint} in a more compact form as
\begin{equation}
   \begin{bmatrix}
   \bar{x}_{0|k}\\
   \mathbf{c}_k
   \end{bmatrix}^\top  \!\!
   W_1\left(K_k\right)
   \begin{bmatrix}
   \bar{x}_{0|k}\\
   \mathbf{c}_k
   \end{bmatrix}
   +\frac{\gamma}{1-\gamma}\tr\bigl( \Omega\tP_k \bigr)
   \leq \varepsilon_k , \label{eqn:constraint:compact form involving (x,c)}
   \end{equation}
where $W_1(K_k)$ is a function of $K_k$ defined as \begin{equation}
W_1\left(K_k\right):=
\begin{bmatrix}
C^\top C+M_x^\top HM_x&M_x^\top HM_c\\
M_c^\top HM_x&M_c^\top HM_c
\end{bmatrix},  \label{eqn:constraint:weighting matrix W1}
\end{equation}
and
$ H:=\text{diag}(\gamma C^\top \!C,\ldots,\gamma^{N-1} C^\top \!C,\gamma^N \!\tP_k) \succcurlyeq 0$.

To summarise, the MPC optimisation solved at time $k$ is
\begin{equation}
 J^\ast (x_{k}, K_k)\!:=\! \min_{\mathbf{c}_k}\{J({x}_{k}, K_k, \mathbf{c}_k) ~|~
\eqref{eqn:constraint:compact form involving (x,c)} ~\text{with}~ \bar{x}_{0|k}\!=\!x_k \}, \label{eq:mpc_optimisation}
\end{equation}
and its solution for any feasible $x_{k}$, $K_k$ and $\varepsilon_k$ is denoted $\mathbf{c}^*_k\left(x_{k}, K_k, \varepsilon_k \right)$.
For simplicity we write this solution as $\mathbf{c}^*_k$, with the understanding that this vector depends on $x_k$, $K_k$ and $\varepsilon_k$. The corresponding predicted nominal state trajectory is given by
\begin{align}
  &\bar{x}^*_{i|k} = \Phi_k^i x_{k}+\sum_{j=0}^{i-1} \Phi_k^{i-1-j}Bc^*_{j|k} ,
  &
    i &= 1,\ldots,N,
        \label{eqn:open loop system: optimal state trajectory, first mode}\\
  &\bar{x}^*_{N+i|k} = \Phi_k^i\bar{x}^*_{N|k},
  &
    i &= 1,2,\ldots .
        \label{eqn:open loop system: optimal state trajectory, second mode}
\end{align}
The MPC law at time $k$ is defined by
\begin{equation}
u_k := K_kx_k+c^\ast_{0|k} ,  \label{eqn:closed loop control law}
\end{equation}
and the closed loop system dynamics are given by
\begin{equation}
  x_{k+1}=\Phi_k x_k+Bc^\ast_{0|k} (x_k, K_k, \varepsilon_k)+\omega_k ,
  \label{eqn:closed loop system dynamics, using the first element of optimal input sequence}
\end{equation}
where $\omega_k$ is the disturbance realisation at time $k$.

In the remainder of this paper, we discuss how to choose $\varepsilon_k$ so as to guarantee recursive feasibility in Section~\ref{section:recursive feasibility};
how to generate a set of feedback gains with desirable properties offline in Section~\ref{section:multiobjective optimisation and dynamic programming};
how to select $K_k$ from among this set of candidates with the aim of minimising predicted costs while retaining the recursive feasibility guarantee of online MPC optimisations in Section \ref{section:dynamic selection};
and how the choices of $P_k$ and $\tP_k$ given in \eqref{eqn:lyap P_k} and \eqref{eq:Plyap} allow for a guarantee of quadratic stability and satisfy constraint \eqref{eqn:constraint:probability constraint of the original form} respectively under the MPC law \eqref{eqn:closed loop control law} in Section \ref{section:smpc algorithm} and Section \ref{section:constraint satisfaction}.

\section{Recursive Feasibility} \label{section:recursive feasibility}
Recursively feasible MPC strategies have the property that the MPC optimisation problem is guaranteed to be feasible at every time step if it is initially feasible. This property is typically ensured by imposing a terminal constraint that requires the predicted  system state to lie in a particular set at the end of the prediction horizon \cite{kouvaritakis2015mpcbook}. For a deterministic MPC problem, if an optimal solution can be found at current time, then the \textit{tail sequence}, namely the optimal control sequence shifted by one time step, will be a feasible but suboptimal solution at the next time instant if the terminal constraint is defined in terms of a suitable invariant set for the predicted system state \cite{Kerrigan2000}. For a robust MPC problem with bounded additive disturbances, recursive feasibility can likewise be guaranteed by imposing a terminal constraint set that is robustly invariant. However, this approach is not generally applicable to systems with unbounded additive disturbances, and in general it is not possible to ensure recursive feasibility in this context while guaranteeing constraint satisfaction at every time instant.

In this section we propose a method for guaranteeing recursive feasibility of the MPC optimisation, which does \emph{not} rely on terminal constraints.
Instead recursive feasibility is ensured, despite the presence of unbounded disturbances, by allowing the constraint on the discounted sum of \textit{violation probabilities} to be time-varying. For every time step $k > 0$, the approach uses the optimal sequence computed at time $k-1$ to determine a value of $\varepsilon_{k}$ that is necessarily feasible at time $k$.

We use the notation $\widetilde{\mathbf{c}}_{k+1}$ to denote a time-shifted version of $\mathbf{c}^*_k$, defined by
\begin{equation}
\widetilde{\mathbf{c}}_{k+1}:=E\mathbf{c}^*_k, \label{eqn:definition: time shifted c sequence}
\end{equation}
where $E$ is the matrix such that $E \mathbf{c}=[c_1^\top ,\ldots,c_{N-1}^\top ,\mathbf{0}]^\top $ if $\mathbf{c}=[c_0^\top ,\ldots,c_{N-1}^\top ]^\top $.

\begin{lemma} \label{theorem:theorem for recursive feasibility}
\blue{Given initial feasibility at time $k=0$}, the MPC optimisation~\eqref{eq:mpc_optimisation} is recursively feasible if $\varepsilon_k$ is defined at each time $k=1,2,\ldots$ as
\begin{equation}
  \varepsilon_{k}:=
     \begin{bmatrix}
   x_{k}\\
   \widetilde{\mathbf{c}}_k
   \end{bmatrix}^\top  \!\!
   W_1\left(K_{k-1}\right)
   \begin{bmatrix}
   x_{k}\\
   \widetilde{\mathbf{c}}_k
   \end{bmatrix}
   +\frac{\gamma}{1-\gamma}\tr\bigl( \Omega\tP_{k-1} \bigr)  .
    \label{eqn: updating epsilon: optimisation by solving which gives new epsilon}
\end{equation}
\end{lemma}
Equation \eqref{eqn: updating epsilon: optimisation by solving which gives new epsilon} provides an explicit expression for $\varepsilon_{k}$ for all $k > 0$ in terms of \eqref{eqn:open loop system: optimal state trajectory, first mode}, \eqref{eqn:open loop system: optimal state trajectory, second mode} and disturbance realisations as
\begin{multline}
\hspace{-2.5mm} \varepsilon_k = \sum_{i=0}^{N-1} \gamma^i \bigl\| C \bigl(\bar{x}_{i+1|k-1}^* + \Phi^i_{k-1}\omega_{k-1} \bigr)   \bigr\|^2 \\
+ \gamma^N \bigl\| \bar{x}_{N+1|k-1}^*+ \Phi^N_{k-1}\omega_{k-1}  \bigr\|^2_{\tP_{k-1}}
+ \frac{\gamma}{1- \gamma } \tr ( \Omega \tP_{k-1} ).
\label{eqn:epsilon of next time instant:expression of new epsilon using results obtained at current time}
\end{multline}
Here
\begin{equation} \label{eqn:feasible state sequence}
\bar{x}_{i+1|k-1}^* + \Phi^i_{k-1}\omega_{k-1}:=\bar{x}_{i|k},\quad i=0,\ldots,N
\end{equation}
defines a feasible nominal state sequence predicted at time $k$, obtained by setting $\bar{x}_{0|k} = x_{k}$ and $\bar{u}_{i|k} = K_{k-1} \bar{x}_{i|k}+ \widetilde{c}_{i|k}$. Note that in constructing this feasible sequence, we still use $K_{k-1}$ and therefore $\tP_{k-1}$ as the corresponding terminal matrix, and that the feedback gain is updated after the update of $\varepsilon_k$.

Essentially, the optimisation problem to be solved at each time step is feasible because the parameter $\varepsilon_k$ is updated via~\eqref{eqn:epsilon of next time instant:expression of new epsilon using results obtained at current time} using knowledge of the disturbance $w_{k-1}$ obtained from the measurement of the current state $x_k$. In this respect, the approach is similar to constraint-tightening methods that have previously been applied in the context of stochastic MPC (e.g.~\cite{korda14,Lorenzen17ieeetr,fleming17}) in order to ensure recursive feasibility and constraint satisfaction in closed loop operation. However, each of these methods requires that the disturbances affecting the controlled system be bounded, and they become more conservative as the degree of conservativeness of the assumed disturbance bounds increases. The approach proposed here avoids this requirement and instead ensures closed loop constraint satisfaction using the analysis of sequence $\{\varepsilon_k\}_{k=0}^{\infty}$ (as will be detailed in Section~\ref{section:constraint satisfaction}).

The key to this method lies in the definition of the vector $\widetilde{\mathbf{c}}_{k+1}$. If this vector were optimised, rather than defined by the suboptimal control input~\eqref{eqn:definition: time shifted c sequence}, then it would be possible to reduce the MPC cost~\eqref{eq:mpc_optimisation}. However this would require more computational effort than is needed to evaluate~\eqref{eqn:epsilon of next time instant:expression of new epsilon using results obtained at current time} and lose the guarantee of satisfying \eqref{eqn:constraint:probability constraint of the original form} in closed loop.

\section{Multiobjective Optimisation and Dynamic Programming} \label{section:multiobjective optimisation and dynamic programming}
In this section, we generate a set of strictly stabilising feedback gains offline, from which $K_k$ in \eqref{eqn:nominal control input:first mode}-\eqref{eqn:nominal control input:second mode} is selected online.
Considering the cost function in \eqref{eqn:cost:original form of the cost function} and the constraint \eqref{eqn:constraint:probability constraint of the original form} as conflicting objectives,
we formulate a set of multiobjective optimisation problems whose solutions provide feedback gains representing a trade-off between these two objectives.
These multiobjective optimisation problems are written using linear scalarisation \cite{Pistikopoulos2016MOP} for given $x_0\in\mathbb{R}^{n_x}$ in the form
\begin{align} \tag{$\mathcal{P}_1$}
&\!\min_{u_0,u_1,\ldots}(1-\mu)\!\sum_{i=0}^{\infty} \gamma^i \mathbb{E} \big\{\norm{Cx_i}^2\big\} \!+\! \mu\!\sum_{i=0}^{\infty} \mathbb{E} \big\{ \norm{ x_i }^2_Q \!+\! \norm{ u_i }^2_R  \big\} \label{eqn:DP cost:original form with expectation}\\
&~\text{s.t.~ } x_{i+1}=Ax_i+Bu_i+\omega_i,~i=0,1,\ldots \label{eq:actual dynamics}
\end{align}
where $\mu \in (0,1]$ is a weighting parameter. Here $\omega_i$ is an i.i.d. random variable with the same statistics as given in \eqref{eq:disturbance statistics}.
Note that in \eqref{eqn:DP cost:original form with expectation} the first part of the objective is an
approximation of the LHS of the constraint \eqref{eqn:constraint:probability constraint of the original form} via Chebyshev's inequality and the second part has the same form as the cost function in problem \eqref{eqn:cost:original form of the cost function}. Possible trade-offs between these competing objectives can be explored using different values of $\mu$, and their solutions provide a set of strictly stabilising feedback gains. We denote this set of feedback gains as $\mathcal{K}$.

We propose to solve \eqref{eqn:DP cost:original form with expectation} by Dynamic Programming \cite{bellman1966dynamic}
via its equivalent deterministic counterpart in the form of
\begin{align}
&\hspace{-2mm}\min_{\bar{u}_0,\bar{u}_1,\ldots}( 1-\mu )\sum_{i=0}^{\infty} \gamma^i \norm{C\bar{x}_i}^2+ \mu \sum_{i=0}^{\infty} ( \norm{ \bar{x}_i }^2_Q + \norm{ \bar{u}_i }^2_R ) \label{eqn:DP cost:deterministic form} \tag{$\mathcal{P}_2$} \\
&\hspace{-2mm}~\text{s.t. } \bar{x}_{i+1}=A\bar{x}_i+B\bar{u}_i,~i=0,1,\ldots  \label{eqn:nominal dynamics}
\end{align}
where $\bar{x}_i$ and $\bar{u}_i$ are mean values of $x_i$ and $u_i$ respectively, with $\bar{x}_0=x_0$.
From the quadratic form of the objective function, the optimal solution to \eqref{eqn:DP cost:deterministic form} is a linear feedback control law, and its corresponding feedback gain has desirable properties as will be shown in Theorem~\ref{theorem:desirable properties of DP iteration}. In the first instance, we show the equivalence between \eqref{eqn:DP cost:original form with expectation} and \eqref{eqn:DP cost:deterministic form} in Theorem~\ref{theorem:equivalence between DP costs}, \blue{which is in agreement with the certainty equivalence principle \cite[Section 8.4]{Astrom06Stochastic}.}
\begin{theorem} \label{theorem:equivalence between DP costs}
Problems \eqref{eqn:DP cost:original form with expectation} and \eqref{eqn:DP cost:deterministic form} are equivalent in the sense that their optimal solutions are linear feedback control laws with the same feedback gains.
\end{theorem}
\begin{IEEEproof}
We begin by deriving the solution of \eqref{eqn:DP cost:original form with expectation}. Let
\begin{align*}
&(1-\mu)\bar{J}_0(x_0) +\mu \hat{J}_0(x_0) := \\
&\!
\lim_{T \!\to \infty} \min_{u_0,\ldots,u_{T\!-\!1}}\!\!\!\EE \bigl\{(1\!-\!\mu)\!\sum_{i=0}^{T-1}\! \gamma^i \norm{Cx_i}^2 \!\!+\! \mu\!\sum_{i=0}^{T-1} ( \norm{ x_i }^2_Q \!\!+\!\! \norm{ u_i }^2_R)  \bigr\}\\
&\hspace{1cm}\text{s.t. \eqref{eq:actual dynamics},}
\end{align*}
where $\bar{J}_0(x_0)$ and $\hat{J}_0(x_0)$ \blue{are optimal cost-to-go functions} denoting the optimal values of
$\lim_{T\to\infty}\sum_{i=0}^{T-1} \gamma^i \EE\{\norm{Cx_i}^2\} $ and $\lim_{T\to\infty}\sum_{i=0}^{T-1} \EE\{ \norm{ x_i }^2_Q \!+\! \norm{ u_i }^2_R\}$, respectively, for the initial state $x_0$. \blue{Here $u_i$, $i \geq 0$ and $x_i$, $i\geq 1$ are decision variables corresponding to the control inputs and states of the system \eqref{eq:actual dynamics}. By solving the Bellman equations \cite{bellman1966dynamic} backwards in time with initial conditions $\bar{J}_T(x_T)=0$ and $\hat{J}_T(x_T)=0$ where $\bar{J}_T(x_T)$, $\hat{J}_T(x_T)$ are the optimal cost-to-go functions starting from state $x_T$, it can be shown that}
$
\bar{J}_0(x_0)=x_0^\top \bar{S}_0 x_0+r_0, ~
\hat{J}_0(x_0)=x_0^\top \hat{S}_0 x_0+v_0.
$
\blue{Here $\bar{S}_0$, $\hat{S}_0$ are positive semidefinite matrices and $r_0$, $v_0$ respectively represent discounted and undiscounted sums of variances of $x_i$ for $i=1,\ldots,T$.}
Due to the quadratic form of the objective function and the assumption of zero-mean disturbance $\omega_i$, the optimal control input $u_i$ is a linear function of $x_i$ for all $i\geq 0$. Denoting the optimal feedback gain as $H_i$,
we obtain
the following DP iteration for $i=T-1,\ldots, 0$:
\begin{align}
\bar{S}_i &= C^\top C+ \gamma (A+BH_i)^\top  \bar{S}_{i+1} (A+BH_i)     , \label{eqn:DP iteration:Pbar}\\
\hat{S}_i &= Q+H_i^\top RH_i + (A+BH_i)^\top  \hat{S}_{i+1} (A+BH_i)    ,\label{eqn:DP iteration:Phat}\\
r_i &= \gamma \left( r_{i+1} + \tr (\bar{S}_{i+1} \Omega )   \right) ,\label{eqn:DP iteration:r}\\
v_i &= \tr \bigl(\hat{S}_{i+1} \Omega\bigr) + v_{i+1}                          ,\label{eqn:DP iteration:v}
\end{align}
where we choose $\bar{S}_T=\hat{S}_T=\mathbf{0}$ and $r_T=v_T=0$.
Also, the corresponding feedback gains $H_i$ can be computed \blue{by backward induction} as
\[
H_{i} = -
 \big[ \mu R + B^\top  \Delta_{i+1}  B \big]^{-1} B^\top  \Delta_{i+1}  A
\]
for $i=T-1,\ldots, 0$ with $\Delta_{i+1} = (1-\mu)\gamma \bar{S}_{i+1} +\mu \hat{S}_{i+1}$.
Therefore, to obtain $H_i$, $\bar{S}_i$ and $\hat{S}_i$ it is only necessary to perform the iterations in \eqref{eqn:DP iteration:Pbar}-\eqref{eqn:DP iteration:Phat}, since these are independent of $r_i$ and $v_i$.
By considering the limit as $T\to\infty$ and using similar reasoning, it can be shown that $x_0^\top \bigl((1-\mu)\bar{S}_0 + \mu \hat{S}_0\bigr)x_0$ is the optimal objective of problem \eqref{eqn:DP cost:deterministic form} and the corresponding optimal control law is given for all $i\geq 0$ by $\bar{u}_i = H_i \bar{x}_i$.
\end{IEEEproof}
Theorem \ref{theorem:equivalence between DP costs} demonstrates that we only need to solve problem \eqref{eqn:DP cost:deterministic form} to obtain the optimal feedback gains for \eqref{eqn:DP cost:original form with expectation}.
To suit our purposes, we reverse the time indexing in \eqref{eqn:DP iteration:Pbar} and \eqref{eqn:DP iteration:Phat} and define $\bar{P}_i:= \bar{S}_{T-i}$, $\hat{P}_i:=\hat{S}_{T-i}$ for $i=0,\ldots,T$ and $L_i := H_{T-i}$ for $i=1,\ldots,T$.
The resulting DP iteration is
\begin{align}
L_{i+1} &=
-  \big[ \mu R + B^\top \Sigma_i B \big]^{-1}  B^\top  \Sigma_i A, \label{eqn:DP iteration:Ldef}\\
\bar{P}_{i+1} &= C^\top C+ \gamma (A+BL_{i+1})^\top  \bar{P}_i (A+BL_{i+1})     , \label{eqn:DP iteration:reverse Pbar}\\
\hat{P}_{i+1} &= Q+L_{i+1}^\top RL_{i+1} + (A+BL_{i+1})^\top  \hat{P}_i (A+BL_{i+1})    ,\label{eqn:DP iteration:reverse Phat}
\end{align}
with $\Sigma_i := \gamma (1 - \mu) \bar{P}_i  + \mu \hat{P}_i$, for $i=0,\ldots, T-1$. These provide the solution of \eqref{eqn:DP cost:deterministic form} in the limit as $T\to \infty$ \blue{and the optimal value of the objective function of problem \eqref{eqn:DP cost:deterministic form} is $x_0^\top ( (1-\mu)\bar{P}_\infty + \mu \hat{P}_\infty) x_0$ for any given $x_0 \in \mathbb{R}^n$, where $\bar{P}_\infty$ and $\hat{P}_\infty$ are defined in the following theorem.}

\begin{theorem}\label{theorem:desirable properties of DP iteration}
Consider equations \eqref{eqn:DP iteration:Ldef}-\eqref{eqn:DP iteration:reverse Phat}. Under Assumption~\ref{assumption: controllability and observability}: (a) there exist matrices $\bar{P}_{\infty} \succcurlyeq 0$, $\hat{P}_{\infty} \succ 0$ such that for arbitrary positive semidefinite matrices $\bar{P}_0$ and $\hat{P}_0$  we have
\[
\lim_{i \to \infty} \bar{P}_i=\bar{P}_{\infty} ~\text{and}~ \lim_{i \to \infty} \hat{P}_i=\hat{P}_{\infty}   ,
\]
(b) for all $\mu \in (0,1]$ the feedback gain $L_{\infty}$ is strictly stabilising, where
\begin{equation} \label{eq:steady state Linf}
 L_{\infty} :=  - \bigl[ \mu R+B^\top \Sigma_\infty B \big]^{-1} \! B^\top \Sigma_\infty A  
\end{equation}
with $\Sigma_\infty := \gamma (1-\mu) \bar{P}_{\infty} + \mu \hat{P}_{\infty}$, and
(c) the matrices $\bar{P}_{\infty}$ and $\hat{P}_{\infty}$ \blue{satisfy the following Lyapunov matrix equations}:
\begin{align}
\bar{P}_{\infty} \!=&C^\top C + \gamma \left( A+BL_{\infty} \right)^\top  \bar{P}_{\infty} \left( A+BL_{\infty} \right),  \label{eq:steady state Pbar}\\
\hat{P}_{\infty} \!=&Q + L_{\infty}^\top RL_{\infty} + \left( A+BL_{\infty} \right)^\top  \!\hat{P}_{\infty} \left( A+BL_{\infty} \right). \label{eq:steady state Phat}
\end{align}
\end{theorem}
\begin{IEEEproof}
These results can be derived using well-known DP techniques, which can be found, for example, in \cite{Bertsekas:2000}.
\end{IEEEproof}
The solutions $L_{\infty}$, $\bar{P}_{\infty}$, $\hat{P}_{\infty}$ are functions of $\mu$ and are denoted $L_{\infty}(\mu)$, $\bar{P}_{\infty}(\mu)$, $\hat{P}_{\infty}(\mu)$ in the remainder of this paper.
With $\mu=1$, $L_{\infty}(1)=K_{LQ}$, where $K_{LQ}$ is the LQ-optimal feedback gain that minimises the second part of the objective in \eqref{eqn:DP cost:deterministic form}, whereas with $\mu=0$ $L_{\infty}(0) = \!-\! \left( B^\top  \bP_{\infty}(0) B \right)^{\dagger}  B^\top   \bP_{\infty}(0) A$, which is optimal with respect to the first part of the objective. However,
the gain $L_\infty(0)$ may not be stabilising and unique due to discounting and
no penalty on control inputs if $\mu=0$, and we therefore restrict the weighting parameter $\mu$ to the interval $(0,1]$.

\begin{remark} \label{remark:equivalence to a nonconvex problem}
The fixed point $(L_\infty(\mu), \bar{P}_\infty(\mu), \hat{P}_\infty(\mu)\bigr)$ of the iteration \eqref{eqn:DP iteration:Ldef}-\eqref{eqn:DP iteration:reverse Phat} coincides
with the minimising argument of
\begin{align*}
&\min_{G \in \mathbb{R}^{n_u \times n_x},Z_1 \succcurlyeq 0 , Z_2 \succcurlyeq 0 }~\tr( (1-\mu)Z_1+\mu Z_2)\\
&\qquad ~~~~~\text{s.t.}  ~
\begin{aligned}[t]
Z_1 &=C^\top \!C + \gamma \left( A+BG \right)^\top  \!\!Z_1\! \left( A+BG \right), \\
Z_2 &=Q + G^\top \!RG + \left( A+BG\right)^\top  \!\!Z_2\! \left( A+BG \right).
\end{aligned}
\end{align*}
However, the iteration is generally preferred over solving this equivalent problem directly since it is nonconvex in variables $G,Z_1,Z_2$, with no obvious convexifying transformation.
\end{remark}

Next, we give procedures, which are executed of\mbox{}f\mbox{}line, to generate feedback gains based on a sequence of positive weighting parameters for gain selection methods in Section~\ref{section:dynamic selection}.\\

\noindent \textbf{$\mathcal{K}$ Generation} (Offline):
\begin{itemize}
\item[(1)] Choose appropriately a sequence $\{\mu_i\}_{i=1}^m$, with $\mu_i  \in  (0,1]$ in ascending order, and $\mu_m = 1$;
\item[(2)] For each $i=1,\ldots,m$, solve problem \eqref{eqn:DP cost:deterministic form} with $\mu=\mu_i$ by executing iterations
\eqref{eqn:DP iteration:Ldef}-\eqref{eqn:DP iteration:reverse Phat};
\item[(3)] Obtain the set of strictly stabilising feedback gains $\mathcal{K}:=\{L_{\infty}(\mu_i)\}_{i=1}^m$.
\end{itemize}
In the $\mathcal{K}$ Generation, there should be a sufficiently large number of elements in the sequence $\{\mu_i\}_{i=1}^m$ so that $L_{\infty}$ can be adequately approximated on the intervals between consecutive points in this sequence. More importantly, $\mu_1$ should be appropriately chosen close to $0$ while ensuring that $L_{\infty}(\mu_1) \neq L_{\infty}(0)$ and $\bar{P}_\infty(\mu_1) \neq \bar{P}_\infty(0)$ if the latter is strictly stabilising, and hence $L_{\infty}(\mu) \neq L_{\infty}(0)$ $\forall \mu \geq \mu_1$ by the uniqueness of solutions to problem \eqref{eqn:DP cost:deterministic form} and the monotonicity property of $\bP_{\infty}(\mu)$ as will be given in the next section. Furthermore,
step (2) can be warm-started by using $\bar{P}_{\infty}(\mu_i)$ and $\hat{P}_{\infty}(\mu_i)$ to initialise the iteration with  weighting parameter $\mu_{i+1}$, thereby reducing considerably the time required to solve \eqref{eqn:DP cost:deterministic form} for each value of $\mu$. Note also that in step (3) the sets $\{\bar{P}_{\infty}(\mu_i)\}_{i=1}^m$, $\{\hat{P}_{\infty}(\mu_i)\}_{i=1}^m$ are obtained.

\section{Dynamic Feedback Gain Selection}\label{section:dynamic selection}
In this section, we provide two methods for dynamically selecting feedback gains from the set $\mathcal{K}$ discussed in Section~\ref{section:multiobjective optimisation and dynamic programming}. These methods are designed such that the recursive feasibility guarantee of \eqref{eq:mpc_optimisation} and a computationally simple online optimisation are retained.
Both methods determine feedback gains that minimise upper bounds on the optimal predicted cost \eqref{eq:mpc_optimisation} and exploit monotonicity of certain functions.
Method \ref{scheme:less computation} requires less online computation and is equivalent to a binary search,  while guaranteeing
almost sure asymptotic convergence of $K_k$ to $K_{LQ}$.
Method \ref{scheme:probably better} is more intuitive and in many cases gives better closed loop performance over short time horizons, but requires slightly more online computation since it requires the online evaluation of a  function of $\mu\in\{\mu_i \}_{i=1}^m$.

We first derive properties of $\bar{P}_{\infty}(\mu)$ and $\hat{P}_{\infty}(\mu)$ that are exploited by both methods, namely that $\bar{P}_{\infty}(\mu)$ and $\hat{P}_{\infty}(\mu)$
are monotonic in $\mu$,
and that $(1-\mu) \bar{P}_{\infty}(\mu)+\mu\hat{P}_{\infty}(\mu)$ is concave for $\mu\in (0,1)$, which implies Lipschitz continuity of $\bar{P}_{\infty}(\cdot)$ and $\hat{P}_{\infty}(\cdot)$ on $(0,1)$.

\begin{lemma}\label{lemma:monotonicity of Pbar and Phat}
For all $\mu_1,\mu_2$ such that $0 < \mu_1 \leq \mu_2 \leq 1$, $\bar{P}_{\infty}(\cdot)$, $\hat{P}_{\infty}(\cdot)$ satisfy
\begin{align*}
\bar{P}_{\infty}(\mu_1) &\preccurlyeq \bar{P}_{\infty}(\mu_2), \\
\hat{P}_{\infty}(\mu_1) &\succcurlyeq \hat{P}_{\infty}(\mu_2).
\end{align*}
\end{lemma}
\begin{lemma}\label{lemma:concavity}
Let $S(\mu) := (1-\mu)\bar{P}_{\infty}(\mu) + \mu \hat{P}_{\infty}(\mu)$, then $S(\cdot)$ is concave on $(0,1)$.
\end{lemma}
\begin{lemma}\label{lemma:lipschitz continuity}
$\bar{P}_{\infty}(\cdot)$, $\hat{P}_{\infty}(\cdot)$ are Lipschitz continuous on $(0,1)$.
\end{lemma}

\subsection{Gain selection method 1}
This section describes a method for selecting the gain $K_k$ online and determines the properties of the sequence $\{\bar{\mu}_k\}_{k=0}^{\infty}$ generated by the following procedure.


\begin{method}\label{scheme:less computation} At each time step $k=1,2,\ldots$
\begin{itemize}
\item[(1)] Compute
\begin{equation}
\mathbf{c}^o(x_k) :=
   \arg\min_{\mathbf{c}}
   \begin{bmatrix}
   x_k\\
   \mathbf{c}
   \end{bmatrix}^\top  \!\!
   W_1\left(K_{k-1}\right)
   \begin{bmatrix}
   x_k\\
   \mathbf{c}
   \end{bmatrix} \label{eqn:min in scheme 1 to make space for K}
\end{equation}
where $W_1(K_{k-1})$ is defined in \eqref{eqn:constraint:weighting matrix W1};
\item[(2)] Compute
\begin{align}
\bar{\mu}_k &:=\max \big\{\!
   \arg\max_{\mu \in \{\mu_i\}^m_{i=1}}~\frac{\gamma}{1-\gamma}\tr( \Omega\bar{P}_{\infty}(\mu) ) \nonumber \\
   &\quad~~ \text{s.t. } \nonumber \\
   & \hspace{-5mm}
   \begin{bmatrix}
   x_k\\
   \!\mathbf{c}^o(x_k)\!
   \end{bmatrix}^\top  \!\!
   W_1\big( K_{k-1} \big)\!\!
   \begin{bmatrix}
   x_k\\
   \!\mathbf{c}^o(x_k)\!
   \end{bmatrix}
   \!+\! \frac{\gamma}{1\!-\!\gamma}\tr( \Omega\bar{P}_{\infty}(\mu) ) \!\leq\! \varepsilon_k \!\big\} \label{eqn: maximisation in scheme 1}
\end{align}
where $\varepsilon_k$, $\bar{P}_{\infty}(\mu)$ are defined in \eqref{eqn: updating epsilon: optimisation by solving which gives new epsilon}, \eqref{eq:steady state Pbar} respectively;
\item[(3)] Set $K_k:=L_{\infty}(\bar{\mu}_k)$, $\tP_k:=\bar{P}_{\infty}(\bar{\mu}_k)$ and $P_k:=\hat{P}_{\infty}(\bar{\mu}_k)$.
\end{itemize}
\end{method}
Step~(1) determines the \textit{perturbation sequence} ${\bf c}^o(x_k)$ that minimises the LHS of constraint~\eqref{eqn:constraint:compact form involving (x,c)} with $K_k=K_{k-1}$ and $\tP_k=\tP_{k-1}$.
Step~(2) then chooses $\bar{\mu}_k$ as the largest element of the sequence $\{\mu_i\}_{i=1}^m$ such that ${\bf c}^o(x_k)$ is feasible for the constraint \eqref{eqn:constraint:compact form involving (x,c)}.
To show this, note that by combining \eqref{eqn:definition: time shifted c sequence}, \eqref{eqn: updating epsilon: optimisation by solving which gives new epsilon} and step (3) we obtain
\begin{multline*}
\varepsilon_k =   \begin{bmatrix}
   x_k\\
   E\mathbf{c}^*_{k-1}
   \end{bmatrix}^\top  \!\!
   W_1\left( L_{\infty}(\bar{\mu}_{k-1}) \right)
   \begin{bmatrix}
   x_k\\
   E\mathbf{c}^*_{k-1}
   \end{bmatrix}  \\
   +\frac{\gamma}{1-\gamma}\tr( \Omega\bar{P}_{\infty}(\bar{\mu}_{k-1}) ) .
\end{multline*}
From Lemma~\ref{lemma:monotonicity of Pbar and Phat},
\eqref{eqn: maximisation in scheme 1} is therefore equivalent to
\begin{subequations}
\begin{align}
& \bar{\mu}_k  :=
   \max_{\mu \in \{\mu_i\}^m_{i=1}}
   \mu
   \label{eqn:scheme 1: optimisation to show main idea} \\
   & \hspace{13.5mm} \text{s.t. } \nonumber \\
   & \frac{\gamma}{1\!-\!\gamma}\tr( \Omega \big( \bar{P}_{\infty}(\mu) \!-\! \bar{P}_{\infty}(\bar{\mu}_{k-1}) \big) )
   \!\leq\!
   \begin{bmatrix}
   x_k\\
   E\mathbf{c}^*_{k-1}
   \end{bmatrix}^\top  \!\!
   \Bigl[
   \star
   \Bigr]
   \begin{bmatrix}
   x_k \\
   E\mathbf{c}^*_{k-1}
   \end{bmatrix} \nonumber
   \\
   & \hspace{33mm}
   -  \begin{bmatrix}
   x_k\\
   \mathbf{c}^o(x_k)
   \end{bmatrix}^\top  \!\!
   \Bigl[
   \star
   \Bigr]
   \begin{bmatrix}
   x_k\\
   \mathbf{c}^o(x_k)
   \end{bmatrix} ,  \label{eq:the key constraint in scheme 1}
\end{align}
\end{subequations}%
where $\star$ denotes $W_1\left( L_{\infty}(\bar{\mu}_{k-1}) \right)$ and the RHS of \eqref{eq:the key constraint in scheme 1} is nonnegative due to the definition of $\mathbf{c}^o(x_k)$ in step (1).
The aim of Method~1 is therefore to use the slack introduced into the constraint through the choice of $\mathbf {c}^o(x_k)$ in step (1) in order to maximise $\bar{\mu}_k$
in step (2) subject to
\[
\begin{bmatrix}
   x_k\\
   \!\mathbf{c}^o(x_k)\!
   \end{bmatrix}^\top  \!\!
   W_1\big( K_{k-1} \big)\!\!
   \begin{bmatrix}
   x_k\\
   \!\mathbf{c}^o(x_k)\!
   \end{bmatrix}
   \!+\! \frac{\gamma}{1\!-\!\gamma}\tr \bigl( \Omega\tP_k \bigr) \!\leq\! \varepsilon_k ,
\]
which implies that a \textit{perturbation sequence} $\mathbf{c}'$ exists satisfying
\begin{equation}
\begin{bmatrix}
   x_k\\
   \!\mathbf{c}^o(x_k)\!
   \end{bmatrix}^\top  \!\!
   W_1\big( K_{k-1} \big)\!\!
   \begin{bmatrix}
   x_k\\
   \!\mathbf{c}^o(x_k)\!
   \end{bmatrix}
   =
   \begin{bmatrix}
   x_k\\
   \!\mathbf{c}'\!
   \end{bmatrix}^\top  \!\!
   W_1\big( K_{k} \big)\!\!
   \begin{bmatrix}
   x_k\\
   \!\mathbf{c}'\!
   \end{bmatrix} \label{eq:feasibility retain}
\end{equation}
so that problem \eqref{eq:mpc_optimisation} remains feasible at time $k$.
Therefore Method \ref{scheme:less computation} retains the recursive feasibility guarantee.
From \eqref{eq:feasibility retain} it also follows that
$J(x_k,K_{k},\mathbf{c}')$ is an upper bound on the optimal value of the cost in~\eqref{eq:mpc_optimisation} at time $k$,
and from \eqref{eqn:scheme 1: optimisation to show main idea} and Lemma~\ref{lemma:monotonicity of Pbar and Phat} it follows that $\bar{\mu}_k$ defined in step (2) minimises the trace of the terminal matrix $P_k = \hat{P}_{\infty}(\bar{\mu}_k)$ in this cost.
%

By Lemma \ref{lemma:monotonicity of Pbar and Phat}, step 2 is simply a binary search to determine the largest $\mu \in \{\mu_i\}_{i=1}^m$ satisfying the constraint in \eqref{eqn: maximisation in scheme 1}.
Since the values of $\tr(\Omega\bar{P}_{\infty}(\mu_i))$ can be calculated offline, Method~\ref{scheme:less computation} can be implemented very efficiently.
Note that the initial value $\bar{\mu}_0$ is not determined by Method~\ref{scheme:less computation}, and we therefore use other means of choosing $\bar{\mu}_0 \in \{\mu_i\}_{i=1}^m$ and $K_0=L_{\infty}(\bar{\mu}_0)$ to make the MPC optimisation \eqref{eq:mpc_optimisation} is initially feasible (assuming such a $\bar{\mu}_0 $ exists); this is discussed in Section~\ref{section:example}, Simulation D.

In the remainder of this section, we analyse the properties of the sequence $\{\bar{\mu}_k\}_{k=0}^{\infty}$ generated by Method \ref{scheme:less computation}.
Since $\bar{\mu}_{k-1}$ is a feasible solution to problem \eqref{eqn: maximisation in scheme 1} at time $k$, which implies $\bar{\mu}_{k-1} \leq \bar{\mu}_k$, and since $\bar{\mu}_k$ is upper bounded by $1$ for all $k$, the sequence $\{\bar{\mu}_k\}_{k=0}^{\infty}$ generated by Method \ref{scheme:less computation} is monotonically non-decreasing and convergent.
To derive a stronger convergence result, we make a simplifying assumption.
\begin{assumption}\label{assumption:can solve continuous optimisation}
The maximisation \eqref{eqn: maximisation in scheme 1} can be solved with the optimisation variable $\mu$ varying continuously in the interval $[\bar{\mu}_0, 1]$ rather than being constrained to the finite set $\{\mu_i\}_{i=1}^{m}$.
\end{assumption}
The purpose of Assumption \ref{assumption:can solve continuous optimisation} is to ensure that $\bar{\mu}_k > \bar{\mu}_{k-1}$ whenever the RHS of \eqref{eq:the key constraint in scheme 1} is positive, and hence that $\bar{\mu}_k$ does not converge to a value less than $1$ as a result of the constraint $\mu\in\{\mu_i\}_{i=1}^{m}$ in step (2).
To analyse asymptotic convergence of $\{\bar{\mu}_k\}_{k=0}^{\infty}$, we consider the RHS of \eqref{eq:the key constraint in scheme 1}. By definition, $W_1(K_{k-1})= W_1\bigl(L_\infty(\bar\mu_{k-1})\bigr)$ is at least positive semidefinite so the minimisation in step 1 is well-defined and $\mathbf{c}^o(x_k)$ can be defined uniquely as
\[
\mathbf{c}^o(x_k)=-\left( W_{cc}(\bar{\mu}_{k-1})\right)^\dagger W_{cx}(\bar{\mu}_{k-1})x_k.
\]
Here $W_{cc}(\mu)$, $W_{cx}(\mu)$ are blocks of $W_1 \bigl( L_{\infty}(\mu) \bigr)$ in the partition \eqref{eqn:constraint:weighting matrix W1} such that $W_{cc} = M_c^\top H M_c$, $W_{cx} = M_c^\top H M_x$, and $W_{cc}^\dagger$ is the pseudoinverse of $W_{cc}$.
Since $x_k = \bar{x}_{1|k-1}^\ast + \omega_{k-1}$, it follows that the RHS of \eqref{eq:the key constraint in scheme 1} is equal to
\begin{equation}\label{eq:RHS(w_(k-1))}
\bigl\|z_{k-1} + \bigl(W_{cc}(\bar\mu_{k-1})\bigr)^\dagger W_{cx}(\bar\mu_{k-1}) \omega_{k-1} \bigr\|^2_{W_{cc}(\bar\mu_{k-1})},
\end{equation}
where $z_{k-1} = E \mathbf{c}_{k-1}^\ast + \bigl(W_{cc}(\bar\mu_{k-1})\bigr)^\dagger W_{cx}(\bar\mu_{k-1}) \bar{x}_{1|k-1}^\ast$.

\begin{lemma}\label{lemma:prob_mubound}
There exist $\delta > 0$ and $p_{\delta} > 0$ such that
\begin{equation}\label{eq:prob_mubound}
\!\inf_{\substack{z\in\mathbb{R}^{Nn_u}\\ \mu\in [\bar{\mu}_0,1]}}
\mathbb{P}\big\{
\bigl\|z + \bigl(W_{cc}(\mu)\bigr)^\dagger W_{cx}(\mu) \omega_{k} \bigr\|^2_{W_{cc}(\mu)} \geq \delta \big\}
\geq p_\delta. \!\!\!
\end{equation}
\end{lemma}

We conclude that $K_k$ converges almost surely to $K_{LQ}$.

\begin{theorem}\label{theorem:almost sure converge}
Under Assumption~\ref{assumption:can solve continuous optimisation}, $\bar{\mu}_k \to 1$ as $k\to\infty$ with probability 1.
\end{theorem}

\subsection{Gain selection method 2}
We first give a monotonicity result that can be used to prove the order-preserving property of Riccati Difference Equations (RDE) \cite{bittanti2012riccati}, which is exploited in Method \ref{scheme:probably better}.
\begin{lemma}\label{lemma:min cost determined by terminal P}
$V(x_0,\check{P}_1) \leq V(x_0,\check{P}_2)$ ~$\forall x_0 \in \mathbb{R}^{n_x}$ ~if ~$\check{P}_1 \preccurlyeq  \check{P}_2$, where
\begin{multline}
V(x_0,\check{P}) := \min_{u_0,\ldots,u_{N-1}} \bigl\{ \widetilde{V}(x_0,\mathbf{u},\check{P})\\
\text{s.t.}~x_{i+1}=Ax_i+Bu_i,~i=0,\ldots,N-1  \bigr\}, \label{eqn: standard finite LQR problem}
\end{multline}
and $\widetilde{V}(x_0,\mathbf{u},\check{P}) :=\sum_{i=0}^{N-1} \bigl( \norm{{x}_i}^2_{Q}+ \norm{u_i}^2_{R} \bigr) + \norm{x_N}^2_{\check{P}\,}$,  $\mathbf{u}:=[u_0^\top ,\ldots,u_{N-1}^\top ]^\top$ for $Q, R, \check{P} \succcurlyeq 0$.
\end{lemma}
\begin{IEEEproof}
Let $\mathbf{u}^*(\check{P}) := \arg\min_{\mathbf{u}}\bigl\{\widetilde{V}(x_0,\mathbf{u},\check{P})~\text{s.t.}~x_{i+1}=Ax_i+Bu_i,~i=0,\ldots,N-1 \bigr\}$, then by the ordering between $\check{P}_1$ and $\check{P}_2$ and optimality, we have
$V(x_0,\check{P}_2) = \widetilde{V}\bigl(x_0,\mathbf{u}^*(\check{P}_2),\check{P}_2\bigr) \geq \widetilde{V}\bigl(x_0,\mathbf{u}^*(\check{P}_2),\check{P}_1\bigr)
\geq \widetilde{V}\bigl(x_0,\mathbf{u}^*(\check{P}_1),\check{P}_1\bigr) = V(x_0,\check{P}_1)$.
\end{IEEEproof}
\begin{remark}
Note that Lemma~\ref{lemma:min cost determined by terminal P} holds if $Q$ and $R$ are time-varying. Let $\check{P}_1^o, \check{P}_2^o$ be the matrices satisfying $V(x_0,\check{P}_1)=:x_0^\top  \check{P}_1^{o} x_0$ and $V(x_0,\check{P}_2)=:x_0^\top  \check{P}_2^o x_0$ for all $x_0 \in \mathbb{R}^{n_x}$, then Lemma~\ref{lemma:min cost determined by terminal P} implies $\check{P}_1^o \preccurlyeq \check{P}_2^o$. Therefore the monotonicity property of RDE can be shown by solving \eqref{eqn: standard finite LQR problem} using DP recursion with $N=1$ sequentially backwards in time. This property was proved in \cite{Bitmead1985,T.Nishimura1966} with the assumption that $B$ has full rank, which is not required here.
\end{remark}

We give Method \ref{scheme:probably better} as follows.
\begin{method}\label{scheme:probably better} At each time step $k=1,2,\ldots$
\begin{itemize}
\item[(1)] For all $\mu \in (0,1]$, define
\begin{equation} \label{eqn:min constraint function over c}
\underline{\mathbf{c}} (x_k,\mu) := \arg \min_{\mathbf{c}}~
   \begin{bmatrix}
   x_k\\
   \mathbf{c}
   \end{bmatrix}^\top  \!\!
   W_1\left(L_{\infty}(\mu)\right)
   \begin{bmatrix}
   x_k\\
   \mathbf{c}
   \end{bmatrix},
\end{equation}
where $W_1$ is defined in \eqref{eqn:constraint:weighting matrix W1};
\item[(2)]Compute
\begin{align}
&\hspace{-2mm}\tilde{\mu}:= \nonumber \\
&\hspace{-3mm}\max\! \big\{\! \! \arg\!\!\max_{\mu \in \{\mu_i\}_{i=1}^{m}}\!\!
\begin{bmatrix}
   x_k\\
  \underline{\mathbf{c}}(x_k,\mu)
   \end{bmatrix}^\top  \!\!
   W_1\left(L_{\infty}(\mu)\right)\!
   \begin{bmatrix}
   x_k\\
   \underline{\mathbf{c}}(x_k,\mu)
   \end{bmatrix}
   \!\!+\!
   (\star) \nonumber \\
&\hspace{-2mm}   \text{s.t.}\!\begin{bmatrix}
   x_k\\
   \underline{\mathbf{c}}(x_k,\mu)
   \end{bmatrix}^\top  \!\!
   W_1\left(L_{\infty}(\mu)\right)\!
   \begin{bmatrix}
   x_k\\
   \underline{\mathbf{c}}(x_k,\mu)
   \end{bmatrix}
   \!\!+\!
   (\star)
   \leq \varepsilon_k \!\big\}, \label{eqn:compute mu_tilde in scheme 2}
\end{align}%
where $\star$ here denotes $\frac{\gamma}{1-\gamma}\tr( \Omega\bar{P}_{\infty}(\mu) )$;
\item[(3)]Compute
\begin{align}
&\bar{\mu}_k:= \nonumber \\
&\hspace{-1mm}\max \! \big\{ \! \arg \! \min_{   \mu \in \{\mu_i\}_{i=1}^{m} }\!\! \begin{bmatrix}
   x_k\\
   \underline{\mathbf{c}}(x_k,\mu)
   \end{bmatrix}^\top  \!\!
   W_2\left(L_{\infty}(\mu)\right)\!\!
   \begin{bmatrix}
   x_k\\
   \underline{\mathbf{c}}(x_k,\mu)
   \end{bmatrix} \nonumber \\
&\qquad  \text{s.t. } \tilde{\mu} \geq \mu \geq \bar{\mu}_{k-1} \big\},
\label{eqn:min cost in scheme 2}
\end{align}
where $W_2$ is defined in \eqref{eqn:def W2};
\item[(4)] Set $K_k:=L_{\infty}(\bar{\mu}_k)$,
$\tP_k:=\bar{P}_{\infty}(\bar{\mu}_k)$ and $P_k:=\hat{P}_{\infty}(\bar{\mu}_k)$.
\end{itemize}
\end{method}
Comparing step (1) of Methods \ref{scheme:less computation} and \ref{scheme:probably better},
here $L_{\infty}(\mu)$ replaces $K_{k-1}$ and
$\underline{\mathbf{c}}(x_k,\mu)$ is a function of both $x_k$ and $\mu$. Recalling the equivalence between \eqref{eqn:constraint: short expression of deterministic online constraint} and \eqref{eqn:constraint:compact form involving (x,c)}, we have that
\begin{equation}
   \begin{bmatrix}
   x_k\\
   \mathbf{c}
   \end{bmatrix}^\top  \!\!
   W_1\left(L_{\infty}(\mu)\right)\!
   \begin{bmatrix}
   x_k\\
   \mathbf{c}
   \end{bmatrix}
\!\!=\!\! \sum^{N-1}_{i=0}\gamma^i \norm{ C\bar{x}_{i|k}}^2 \!+ \gamma^N \norm{\bar{x}_{N|k}}^2_{ \bar{P}_{\infty}(\mu)}, \nonumber
\end{equation}%
where $\bar{x}_{i|k}$ evolves according to
$\mathbf{\bar{x}}_k = M_x\left( (L_{\infty}(\mu) \right) x_{k}+M_c\left( (L_{\infty}(\mu) \right) \mathbf{c}$.
Therefore, by Lemmas \ref{lemma:monotonicity of Pbar and Phat} and \ref{lemma:min cost determined by terminal P}, the objective in the maximisation \eqref{eqn:compute mu_tilde in scheme 2} is monotonic in $\mu$ and step (2) can be performed as a binary search. The constraint in \eqref{eqn:compute mu_tilde in scheme 2} makes the constraint \eqref{eqn:constraint:compact form involving (x,c)} as tight as possible with $K_k = L_\infty(\tilde{\mu})$, and the constraints of \eqref{eqn:min cost in scheme 2} imply that \eqref{eq:mpc_optimisation} is
recursively feasible
with Method \ref{scheme:probably better}.
Note that the initial value $\bar{\mu}_0$ is not determined by Method~\ref{scheme:probably better}. Therefore we choose $\bar{\mu}_0 \in \{\mu_i\}_{i=1}^m$ and $K_0=L_{\infty}(\bar{\mu}_0)$ to make the MPC optimisation~\eqref{eq:mpc_optimisation} initially feasible (if such $\bar{\mu}_0$ exists), as discussed in Section~\ref{section:example}, Simulation~D.

Comparing with Method \ref{scheme:less computation},
both terms appearing on the LHS of the constraint in problem \eqref{eqn:compute mu_tilde in scheme 2} are monotonic in $\mu$,
whereas only second term on the LHS of the constraint in problem \eqref{eqn: maximisation in scheme 1}  increases with $\mu$ and the 
first term is fixed. It follows that $\bar{\mu}_k$ obtained by Method \ref{scheme:less computation} is necessarily greater than or equal to $\tilde{\mu}$ defined in \eqref{eqn:compute mu_tilde in scheme 2}, and hence also necessarily
greater than or equal to $\bar{\mu}_k$ obtained by Method \ref{scheme:probably better}.
The sequence $\{\bar{\mu}_k\}_{k=0}^{\infty}$ generated by Method \ref{scheme:less computation} is therefore likely to converge more quickly to 1. On the other hand,
problem \eqref{eqn:min cost in scheme 2} results in a smaller upper bound on the optimal predicted cost \eqref{eq:mpc_optimisation} and hence Method \ref{scheme:probably better} is likely to provide better closed loop performance over the time period required for the sequence $\{K_k\}_{k=0}^\infty$ determined by Method \ref{scheme:less computation} to converge to $K_{LQ}$. 
Over a longer time interval however, Method \ref{scheme:less computation} is likely to perform better due to its earlier convergence to the LQ-optimal feedback gain $K_{LQ}$. These observations are supported by the numerical example in Section \ref{section:example}.

Based on this discussion, we provide the following guideline for choosing the online gain selection method. If the discount factor $\gamma$ in \eqref{eqn:constraint:probability constraint of the original form} is close to $1$, indicating that performance over a long horizon is a priority,
then Method \ref{scheme:less computation} should be chosen. Alternatively if $\gamma$ is close to $0$ and performance in the immediate future is important, then Method \ref{scheme:probably better} is preferable.
Finally, we note that Method \ref{scheme:probably better} requires more computation than Method \ref{scheme:less computation} since \eqref{eqn:min cost in scheme 2} (the objective of which is not necessarily monotonic in $\mu$)
requires an additional search over $\{\mu_i\}_{i=1}^m$.
\begin{remark}
Since $\bar{\mu}_{k-1}$ is a feasible solution to the optimisation problem \eqref{eqn:min cost in scheme 2} at time $k$ and $\bar{\mu}_k$ is upper bounded by $1$ for all $k$, the sequence $\{\bar{\mu}_k\}_{k=0}^{\infty}$ generated by Method \ref{scheme:probably better} is monotonically non-decreasing and convergent.
\end{remark}

Several factors influence the implementation of online gain selection methods. The choice of the sequence $\{\mu_i\}_{i=1}^m$ can affect convergence of $\{\bar{\mu}_k\}_{k=0}^{\infty}$ and hence closed loop performance, since large gaps between successive elements of the sequence $\{\mu_i\}_{i=1}^m$ reduce the likelihood of convergence $\bar{\mu}_k \to 1$.
Therefore it is desirable to choose $m$ to be as large as possible, subject to offline computation and online storage constraints.
To ensure a large feasible set and to steer the closed loop system away from feedback gains $K_k$ that give worse closed loop performance, it is desirable to choose $\mu_1$ close to $0$
with $\mu_{i+1}-\mu_i$  increasing for larger values of $i$.
Alternatively, if the initial conditions of the MPC problem are known offline when the set $\mathcal{K}$ is generated, then it is obviously advantageous to set $\mu_1$ equal to a value $\bar{\mu}_0$ that makes the MPC optimisation initially feasible.

\section{SMPC Algorithm and Stability Condition}\label{section:smpc algorithm}
For deterministic MPC, it can be shown by using the \textit{tail sequence} that, with an appropriate terminal weighting matrix \cite{Mayne2000auto}, optimal MPC predicted costs are monotonically non-increasing along closed loop trajectories. This property does not generally hold in the presence of unbounded disturbances, and in fact the optimal MPC predicted cost defined by \eqref{eq:mpc_optimisation} is not necessarily monotonically non-increasing if $\varepsilon_k$ is defined by \eqref{eqn: updating epsilon: optimisation by solving which gives new epsilon}. However, in this section we show that the proposed approach based on \eqref{eqn:definition: time shifted c sequence}  ensures a closed loop stability bound.
We first state the online MPC algorithm based on the optimisation problem defined in~\eqref{eq:mpc_optimisation}.
\begin{algorithm}(SMPC Algorithm)\label{algorithm:SPMC algorithm}
  At each time-step $k=0,1,\ldots$~:
  \begin{enumerate}[(i).]
  \item
    Measure $x_k$;
  \item If $k>0$, compute $\varepsilon_k$ using \eqref{eqn: updating epsilon: optimisation by solving which gives new epsilon}
and determine $K_k$ using Method \ref{scheme:less computation} or \ref{scheme:probably better};
  \item Solve the quadratically constrained quadratic programming (QCQP) problem \eqref{eq:mpc_optimisation};
  \item Apply the control law \eqref{eqn:closed loop control law}.
  \end{enumerate}
\end{algorithm}
We choose $\varepsilon_0=e$ in step (ii) as will be explained in Section~\ref{section:constraint satisfaction}.
In step (iii), the MPC optimisation can be solved efficiently since there is only one quadratic constraint in \eqref{eq:mpc_optimisation}, for example using a second-order conic program (SOCP) solver or using the algorithm proposed in \cite{Kouvaritakis2002WhoNQ}, which is based on the Newton-Raphson method.
\begin{theorem} \label{theorem:stability result}
Given initial feasibility of problem \eqref{eq:mpc_optimisation} at time $k=0$, by using Algorithm~\ref{algorithm:SPMC algorithm}, the closed loop system satisfies the quadratic stability condition
\begin{align}
\lim_{T\to\infty}\frac{1}{T}\sum_{k=0}^{T-1}\mathbb{E}\{\norm{{x}_{k}}^2_Q\!+\!\norm{u_{k}}^2_R\}
 &\leq  \lim_{T\to\infty}\frac{1}{T}\sum_{k=0}^{T-1} \tr\bigl( \Omega \mathbb{E}\{P_k\}\bigr) \nonumber\\
 &\leq \tr \bigl(\Omega P_0 \bigr) \label{eqn:quadratic stability result}
\end{align}
provided $P_k$ satisfies \eqref{eqn:lyap P_k}.
\end{theorem}

Stability is the overriding requirement and in most MPC literature the optimal predicted cost is chosen as a Lyapunov function suitable for analysing closed loop stability~\cite{Mayne2000auto}. Theorem~\ref{theorem:stability result} is proved via cost comparison, building a connection between the cost of a feasible solution at time $k+1$ and the optimal cost value at time $k$. Similar asymptotic bounds on the time average of a quadratic expected stage cost are obtained in \cite{KOUVARITAKIS2010auto,Cannon09ieetr} and \cite{Chatterjee15Sstability}.
However, in the current context, Theorem~\ref{theorem:stability result} demonstrates that an MPC algorithm can ensure closed loop stability \emph{without} imposing terminal constraints based on an invariant set.

\section{The Behaviour of The Sequence $\{\varepsilon_k\}_{k=0}^{\infty}$ and Constraint Satisfaction }\label{section:constraint satisfaction}
This section considers the properties of the sequence $\{\varepsilon_k\}_{k=0}^{\infty}$ in closed loop operation under Algorithm~\ref{algorithm:SPMC algorithm}.
We first derive a recurrence equation relating the expected value of $\varepsilon_{k+1}$ to $x_k$ and $\varepsilon_k$ by using the explicit expression for $\varepsilon_k$ in~\eqref{eqn:epsilon of next time instant:expression of new epsilon using results obtained at current time}. This allows an upper bound to be determined for the sum of discounted \textit{violation probabilities} on the LHS of~\eqref{eqn:constraint:probability constraint of the original form}. Then with this bound we show that the closed loop system under the control law of Algorithm~\ref{algorithm:SPMC algorithm} satisfies the chance constraint~\eqref{eqn:constraint:probability constraint of the original form} if $\varepsilon_k$ is initialised with $\varepsilon_0=e$.

The following result gives the relationship between  $\varepsilon_k$ and the expected value of $\varepsilon_{k+1}$ for the closed loop system.
\begin{lemma}\label{thm:epsilon}
Given initial feasibility of problem \eqref{eq:mpc_optimisation} at time $k=0$,
if $\varepsilon_k$ is defined by~\eqref{eqn: updating epsilon: optimisation by solving which gives new epsilon} at all times $k\geq 1$ and $\tP_k$ satisfies \eqref{eq:Plyap} for all $k \geq 0$, then in closed loop operation under Algorithm~\ref{algorithm:SPMC algorithm} we have \begin{equation}
    \gamma \Ex[k]{\varepsilon_{k+1}} \leq \varepsilon_k - \norm{Cx_k}^2,~  \forall k \geq 0. \label{eq:epsilon}
  \end{equation}
\end{lemma}

The main result of this section is given next.
\begin{theorem} \label{theorem:closedloop_cc}
  The closed loop system under Algorithm~\ref{algorithm:SPMC algorithm} satisfies the constraint \eqref{eqn:constraint:probability constraint of the original form} if $\varepsilon_0=e$ and problem \eqref{eq:mpc_optimisation} is initially feasible at time $k=0$.
\end{theorem}
\begin{IEEEproof}
The proof of Theorem \ref{theorem:closedloop_cc} is similar to the proof of Theorem 6 in \cite{Yan_discountsmpc_18} and therefore is omitted here.
\end{IEEEproof}

The presence of the discount factor $\gamma\in(0,1)$ on the LHS of~\eqref{eq:epsilon} implies that the expected value of $\varepsilon_k$ can increase as well as decrease. In fact, for values of $\gamma$ close to zero, a rapid initial growth in $\varepsilon_k$ is to be expected, which is in agreement with the interpretation that the constraint~\eqref{eqn:constraint:probability constraint of the original form} penalises \textit{violation probabilities} to a much lesser extent from the initial time in this case. On the other hand, for values of $\gamma$ close to 1, $\varepsilon_k$ can be expected to decrease initially, implying a greater emphasis on the expected number of violations over some initial horizon.

\section{Numerical Example}\label{section:example}
This section describes a numerical example to illustrate the following points:
(i) the closed loop system \eqref{eqn:closed loop system dynamics, using the first element of optimal input sequence}
satisfies the quadratic stability condition \eqref{eqn:quadratic stability result} and the constraint \eqref{eqn:constraint:probability constraint of the original form} when Algorithm~\ref{algorithm:SPMC algorithm} is used without dynamic gain selection or with either gain selection Method~\ref{scheme:less computation} or Method~\ref{scheme:probably better};
(ii) the degree of conservativeness of Chebyshev's inequality is mitigated by using online gain selection procedures in the sense that the long-run expected average costs are improved and the observed constraint violation rates are closer to the imposed limit;
(iii)~gain selection Method~\ref{scheme:probably better} provides better closed loop performance over short time intervals than  Method~\ref{scheme:less computation};
(iv) $K_k$ converges to $K_{LQ}$ with high probability for Method~\ref{scheme:less computation};
(v) the feasible set of initial conditions is enlarged with an appropriate initial feedback gain.
We also \blue{conduct simulations to compare Algorithm \ref{algorithm:SPMC algorithm} with the existing method proposed in \cite{schildbach2014scenario} and }discuss the computation times required by Algorithm~1 and the gain selection methods.

We consider the discrete-time linearised model derived from the continuous-time model of a coupled-tank system \cite{coupledtank} with a sampling interval of $0.05$ sec. This has model parameters
\[
A=\bigl[
\begin{smallmatrix}
0.8207 & 0.04 \\
0.0799 & 0.7808
\end{smallmatrix}\bigr], \quad
B=\bigl[
\begin{smallmatrix}
0.0454 & 0.0011 \\
0.0022 & 0.0443
\end{smallmatrix}\bigr],
\]
and disturbance $\omega_k$ has a multivariate Laplace distribution with zero mean and covariance $\Omega=\Id$. The constraint \eqref{eqn:constraint:probability constraint of the original form} is defined by $\gamma=0.9$, $e=1.5$ and
$C=\bigl[
\begin{smallmatrix}
    0.3 &  0.15\\
    0.1 &  -0.1
\end{smallmatrix}\bigr],
$
and the weighting matrices in the cost of problem \eqref{eqn:cost:original form of the cost function} are given by $Q=R=\Id$. We choose a prediction horizon $N=10$, and the initial value for $\varepsilon_k$ is $\varepsilon_0=e=1.5$. We appropriately choose a sequence $\{\mu_i\}_{i=1}^m$ with $m=290320$, where $0<\mu_i\leq 1$ for all $i$ and $\mu_m=1$. Using this sequence, we generate the set of feedback gains $\mathcal{K}=\{L_{\infty}(\mu_i)\}_{i=1}^m$ offline and store the sets of matrices $\{\bar{P}_{\infty}(\mu_i)\}_{i=1}^m$, $\{\hat{P}_{\infty}(\mu_i)\}_{i=1}^m$. We choose $\bar{\mu}_0=10^{-15}$ so that
\[
K_0=\bigl[
\begin{smallmatrix}
  -18.0749 &  -0.4626\\
   -0.9251 & -17.6123
\end{smallmatrix}\bigr],
\]
and $\Phi_0 = A + BK_0$ is strictly stable.
Simulations A-D all use the initial condition $x_0=[-1,3]^\top$ and the same sequences of disturbances. Note that $K_k=K_{LQ}$ is infeasible for the online MPC optimisation \eqref{eq:mpc_optimisation} at time $k=0$.

\textit{Simulation A} (demonstrating (i) and (ii)):
To estimate empirically the average cost, denoted $J_{\text{average}}$, and the discounted sum of \textit{violation probabilities}, denoted $P_{\text{violation}}$, we run $10^4$ simulations, each of which has a length of $10^4$ time steps, using Algorithm \ref{algorithm:SPMC algorithm} with fixed $K_k=K_0~\forall k\geq0$, with gain selection Method \ref{scheme:less computation}, and with gain selection Method \ref{scheme:probably better}, respectively. We compute the mean value of stage costs over these $10^4$ simulations and count the number of violations in each simulation up to 150 time steps. Simulation results are summarised in Table~\ref{tab:SimAtable}.

\def\colspace{\hspace{1.5ex}}
\begin{table}[h]
\begin{small}
\begin{tabularx}{0.48\textwidth} {
l@{\colspace}|@{\colspace}l@{\colspace}l@{\colspace}l@{\colspace} |  >{\raggedright\arraybackslash}X  }
  & $K_k\!=\!K_0$ & Method \ref{scheme:less computation} & Method \ref{scheme:probably better}&
\\
\hline
$J_{\text{average}}$\rule{0pt}{10pt}
  & $639.9$  & $8.4$  & $96.2$  & $\tr(\Omega P_0)= 640.0$
\\
$P_{\text{violation}}$  & $0.448$  & $0.914$  & $0.769$  & $e=1.5$
\rule{0pt}{10pt}
\end{tabularx}
\end{small}
\caption{Average costs and violation rates for Simulation~A}
\label{tab:SimAtable}
\end{table}

Table~\ref{tab:SimAtable} confirms that the three empirical cost estimates agree with the bound \eqref{eqn:quadratic stability result} and the three estimates for the discounted sum of \textit{violation probabilities} are all smaller than $e=1.5$, implying that constraint \eqref{eqn:constraint:probability constraint of the original form} is satisfied. Note that the cost estimates decrease more slowly as the simulation horizon length continues to increase, and that $\gamma^{150} \approx 1.37 \times 10^{-7}$ so the number of violations occurring at time steps $k>150$ has negligible effect on the estimate of $P_{\text{violation}}$.
Also, cost estimates obtained using the dynamic gain selection methods are considerably smaller than that obtained using a fixed feedback gain. Estimates for $P_{\text{violation}}$ are much closer to the maximum allowed level $e=1.5$ when gain selection methods are used. More specifically, the estimate obtained using gain selection Method \ref{scheme:less computation} is more than double the estimate obtained with a fixed feedback gain. Hence these results show that the conservativeness of Chebyshev's inequality is mitigated by dynamic gain selection.
Moreover, the cost estimate obtained using Method~\ref{scheme:less computation} is smaller than that obtained using Method~\ref{scheme:probably better} since Method \ref{scheme:less computation} achieves convergence of $\bar{\mu}_k$ to $1$ (and hence convergence of $K_k$ to $K_{LQ}$) earlier.
This observation supports the statement in Section \ref{section:dynamic selection} that Method \ref{scheme:less computation} is likely to provide better closed loop performance over longer time periods. \blue{Note that although gain selection methods are employed, there are discrepancies between constraint violation rates observed and the limit imposed. There are four factors that may account for this: (1) Chebyshev's inequality is used to handle the chance constraint, bringing substantial conservativeness; (2)~constraint-tightening is used to ensure recursive feasibility; (3)~$e$ is an upper bound on the discounted sum of \textit{violation probabilities} and there may exist solutions yielding small rates of violation; (4)~the gain selection methods are designed to minimise upper bounds on optimal predicted costs and their impact on increasing constraint violation may be limited.}

\textit{Simulation B} (demonstrating (iii)):
We run $10^4$ simulations, each of which has a length of $40$ time steps, using Algorithm~\ref{algorithm:SPMC algorithm} with fixed $K_k=K_0~\forall k\geq0$, with gain selection Method \ref{scheme:less computation}, and with gain selection Method \ref{scheme:probably better}, respectively.
We compute the mean value of stage costs over these $10^4$ simulations, as shown in Table~\ref{tab:SimBtable}.

\begin{table}[h]
\begin{small}
\begin{tabularx}{0.48\textwidth} {
l @{\colspace} | >{\raggedright\arraybackslash}X
   >{\raggedright\arraybackslash}X
   >{\raggedright\arraybackslash}X}
  & $K_k\!=\!K_0$ & Method \ref{scheme:less computation} & Method \ref{scheme:probably better}
\\
 \hline
$J_{\text{average}}$ \rule{0pt}{10pt}  & $807.2$  & $575.9$  & $500.8$
\end{tabularx}
\end{small}
\caption{Average costs for Simulation B}
\label{tab:SimBtable}
\end{table}

\begin{figure}[t]
  \centering
    \begin{subfigure}[b]{\linewidth}
  \includegraphics[width=\linewidth]{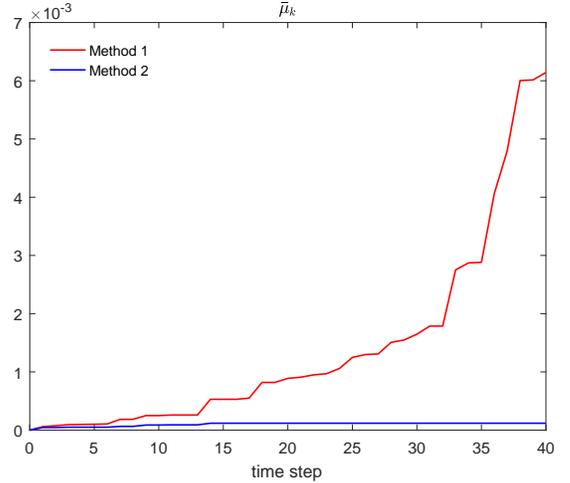}
  \caption{Short simulation horizon}
   \end{subfigure}

  \begin{subfigure}[b]{\linewidth}
  \includegraphics[width=\linewidth]{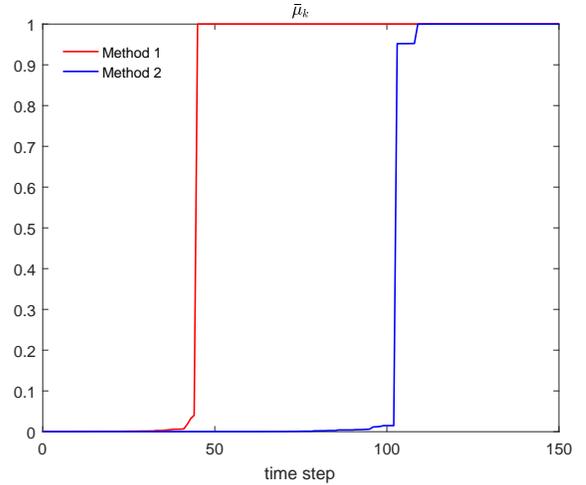}
  \caption{Long simulation horizon}
   \end{subfigure}
   \caption{Evolution of $\bar{\mu}_k$ in Simulation B}
  \label{fig:bar_mu}
  \captionsetup{justification=centering}
\end{figure}

It is clear from Table~\ref{tab:SimBtable} that Method \ref{scheme:probably better} gives a smaller average cost estimate, which is in agreement with the statement in Section \ref{section:dynamic selection} that Method \ref{scheme:probably better} is preferable if short-term performance is prioritised.
This is because $\bar{\mu}_k$ generated by Method \ref{scheme:less computation} generally does not converge to 1 within 40 time steps in this set of simulations, so the average costs are dominated by transient behaviours.
We show an example of sequences $\{\bar{\mu}_k\}_{k=0}^{\infty}$ generated by Methods \ref{scheme:less computation} and~\ref{scheme:probably better} in Figure \ref{fig:bar_mu}, where Figure~\ref{fig:bar_mu}(b) plots the two sequences up to time $k=150$ and Figure~\ref{fig:bar_mu}(a) zooms in on the initial $40$ time steps.
%
These two figures show that $\bar{\mu}_k$ obtained from Method \ref{scheme:less computation} is greater than or equal to that obtained from Method \ref{scheme:probably better} at all times, which is in agreement with the analysis in Section \ref{section:dynamic selection}.

\textit{Simulation C} (demonstrating (iv)):
We run
$10^4$ simulations, each of which has a length of $200$ time steps.
When Algorithm \ref{algorithm:SPMC algorithm} is used with Method~\ref{scheme:less computation}, $\bar{\mu}_{200}=1$ is obtained for every simulation, implying that $K_k$ converges to $K_{LQ}$ in every simulation. On the other hand, when Method \ref{scheme:probably better} is used, the mean value of $\bar{\mu}_{200}$ over the set of $10^4$ simulations is $0.8578$.

\begin{figure}[t]
  \centering
  \includegraphics[width=\linewidth]{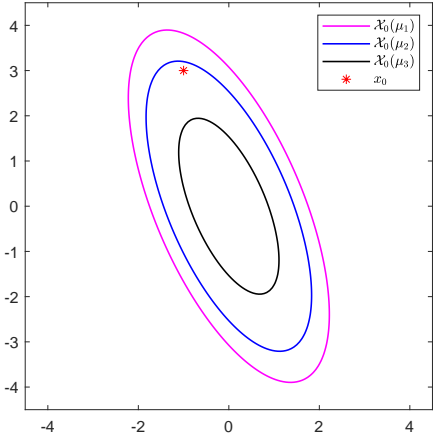}
  \caption{Largest feasible sets of initial conditions}
  \label{fig:set}
\end{figure}

\textit{Simulation D} (demonstrating (v)):
Minimising the LHS of \eqref{eqn:constraint:compact form involving (x,c)} over $\mathbf{c}_k$ yields the largest set of feasible initial conditions,
\begin{multline*}
\mathcal{X}_0(\mu):=\{x : x^\top\bigl(W_{xx}(\mu)-W_{xc}(\mu)W_{cc}^\dagger(\mu) W_{cx}(\mu)\bigr) x \\
+\frac{\gamma}{1-\gamma} \tr (\Omega \bar{P}_{\infty}(\mu)) \leq \varepsilon_0\} ,
\end{multline*}
where $W_{xx}(\mu)$, $W_{xc}(\mu)$, $W_{cx}(\mu)$, $W_{cc}(\mu)$ are the blocks of $W_1 \bigl( L_{\infty}(\mu) \bigr)$ in  \eqref{eqn:constraint:weighting matrix W1}. The feasible set $\mathcal{X}_0$ is plotted for $\mu$ taking values of $\mu_1=10^{-15}$, $\mu_2=10^{-4}$, $\mu_3=2.5 \times 10^{-4}$ in Figure~\ref{fig:set}. Clearly the feasible set is enlarged as $\mu$ is reduced.

\blue{\textit{Simulation E}: To compare the performance of Algorithm \ref{algorithm:SPMC algorithm} with dynamic gain selection with the performance of the scenario-based MPC algorithm in \cite{schildbach2014scenario}, we run $1000$ simulations for each algorithm and each simulation has a length of $200$ time steps.
The system and problem parameters remain the same as initially specified at the beginning of this section, except $N=5$,
$ Q=\bigl[
\begin{smallmatrix}
0&0\\
0&    0.5
\end{smallmatrix}\bigr],~
R=\bigl[\begin{smallmatrix}
1 &    0  \\
0 &    0.01
\end{smallmatrix}\bigr].
$
To implement Algorithm \ref{algorithm:SPMC algorithm} with dynamic gain selection,
$\{L_{\infty}(\mu_i)\}_{i=1}^m$, $\{\bar{P}_{\infty}(\mu_i)\}_{i=1}^m$ and $\{\hat{P}_{\infty}(\mu_i)\}_{i=1}^m$ are computed offline using the sequence $\{\mu_i\}_{i=1}^m$, and $\bar{\mu}_0$ is chosen such that problem \eqref{eq:mpc_optimisation} is feasible at time $k=0$. To implement the algorithm in \cite{schildbach2014scenario}, we choose a violation level $\varepsilon=16.67\%$, which together with the support rank $\rho_1=2$ yields a scenario number $K=11$, and a finite horizon scenario program (FHSCP) is solved at each time step, where the affine disturbance feedback (ADF) policy is used to parameterise control inputs and the FHSCP has $N\,\!K$ $(=\!\!55)$ constraints on predicted states. Then this algorithm ensures that
\[
\sum_{k=0}^{200} \gamma^k \mathbb{P}\left\{ \norm{C x_k} \geq 1 \right\}  \leq e
\]
is satisfied in closed loop.
We compute the mean values of stage costs over these $1000$ simulations and count the number of violations in each simulation.
Simulation results are summarised in Table \ref{tab:comparative}.}
\begin{table}[h]
\begin{small}
\begin{tabularx}{0.48\textwidth} {
l @{\colspace} | >{\raggedright\arraybackslash}X
   >{\raggedright\arraybackslash}X
   >{\raggedright\arraybackslash}X}
  & \cite{schildbach2014scenario} with ADF & Method \ref{scheme:less computation} & Method \ref{scheme:probably better}
\\
 \hline
$J_{\text{average}}$ \rule{0pt}{10pt}  & $62.8651$  & $54.8299$  & $75.2514$
\\
$P_{\text{violation}}$ \rule{0pt}{10pt}  &$1.2971$ &$0.7053$ & $0.6689$    
\end{tabularx}
\end{small}
\caption{Average costs and violation rates for Simulation E}
\label{tab:comparative}
\end{table}

\vspace{-1.2\baselineskip}
\blue{Table \ref{tab:comparative} shows that although Algorithm \ref{algorithm:SPMC algorithm} with dynamic gain selection yields much lower constraint violation rates comparing with the algorithm in \cite{schildbach2014scenario}, Algorithm \ref{algorithm:SPMC algorithm} with gain selection method \ref{scheme:less computation} achieves a better closed loop cost. Also, the implementation of Algorithm \ref{algorithm:SPMC algorithm} has lower computational complexity.}

\textit{Computation times}: Simulations are run in MATLAB R2019a on a computer with
2.20GHz Intel Core i7-8750H CPU and 16GB RAM, and the online MPC optimisation~\eqref{eq:mpc_optimisation} is solved using the root-finding algorithm proposed in \cite{Kouvaritakis2002WhoNQ}. Each online MPC optimisation is solved within 1 millisecond (ms), with an average time of 0.35\,ms. Gain selection Method~\ref{scheme:less computation} requires less than 0.003\,ms at each iteration of Algorithm~\ref{algorithm:SPMC algorithm}, which represents a tiny fraction of the time needed to solve an MPC optimisation.
Gain selection Method \ref{scheme:probably better} requires less than 0.05\,ms at most time steps, which remains a small fraction of the time needed to solve an MPC optimisation. The maximum time observed for implementation of Gain selection Method \ref{scheme:probably better} is 70\,ms.
After $\bar{\mu}_k$ has converged to 1, no gain selection method is executed, so no additional online computation other than the MPC optimisation is required.

\section{Conclusion}\label{section:conclusion}
A stochastic MPC algorithm is proposed that enforces a constraint on
the sum of discounted future \textit{violation probabilities}
and ensures recursive feasibility of the online optimisation
and closed loop constraint satisfaction. Key features are the
design of a constraint-tightening procedure and dynamic gain selection methods, and closed loop
analysis of the tightening parameters. The MPC algorithm
requires knowledge of the first and second moments of the
disturbance and is implemented as a convex QCQP problem.
Stability and constraint satisfaction are ensured for the closed loop system without assuming disturbance bounds.
The results in Sections~III, VI and VII of this paper on recursive feasibility of the MPC optimisation, closed loop stability and constraint satisfaction can be extended to problems with multiple chance constraints,
although solving the corresponding multiobjective optimisation problems to generate stabilising feedback gains for dynamic gain selection becomes more challenging in this case.


\appendices
\section{}
\subsection{Proof of Lemma \ref{lemma: expression for terminal term in constraint}}
Equation \eqref{eq:Plyap} implies that $\tP_k=\sum_{i=0}^{\infty} \gamma^i (\Phi_k^\top)^i C^\top C \Phi_k^i$. Combining this with \eqref{eqn:nominal state:dynamics describing the evolution of nominal state after time N} shows that
\begin{align*}
\sum_{i=N}^{\infty} \!\gamma^i \norm{C \bar{x}_{i|k}}^2
&\!=\!\gamma^N \bar{x}_{N|k}^\top \Bigl( \sum_{j=0}^{\infty} (\gamma^{\frac{1}{2}} \Phi_k^\top )^j  C^\top C (\gamma^{\frac{1}{2}} \Phi_k)^j\Bigr) \bar{x}_{N|k} \nonumber \\
&\!=\!\gamma^N \norm{\bar{x}_{N|k}}^2_{\tP_k}.
\end{align*}
Furthermore, let $\tS_k=\sum_{i=0}^{\infty} \gamma^i X_{i|k}$, then  \eqref{eqn:covariance matrix of state:expression of covariance matrix in term of sums involving W} implies
\begin{multline*}
\hspace{-2mm} \gamma \Phi_k \tS_k \Phi_k^\top  = \sum_{i=0}^{\infty} \gamma^{i+1} \Phi_k X_{i|k} \Phi_k^\top = \sum_{i=0}^{\infty} \gamma^{i+1} \bigl(X_{i+1|k}-\Omega\bigr) \\
=\tS_k - X_{0|k}-\frac{\gamma}{1-\gamma}\Omega,
\end{multline*}
and $\tS_k$ satisfies the Lyapunov equation
\[
\tS_k  = \gamma \Phi_k \tS_k \Phi_k^\top + \frac{\gamma}{1-\gamma}\Omega.
\]
Therefore,
\begin{align*}
&\tr \Bigl(\sum_{i=0}^{\infty} \gamma^i C^\top C X_{i|k} \Bigr)= \tr\bigl(C^\top C \tS_k\bigr) \\
&= \tr\Bigl( C^\top C \sum_{i=0}^{\infty}\gamma^i \Phi_k^i \bigl(\frac{\gamma}{1-\gamma}\Omega\bigr) (\Phi_k^\top)^i \Bigr)
=\frac{\gamma}{1-\gamma} \tr\bigl(\Omega \tP_k\bigr).
\end{align*}
\blue{\subsection{Proof of Lemma \ref{theorem:theorem for recursive feasibility}}
If $K_k=K_{k-1}$ at time $k \geq 1$, then $\widetilde{P}_k=\widetilde{P}_{k-1}$. From equations \eqref{eqn:constraint:compact form involving (x,c)}, \eqref{eqn: updating epsilon: optimisation by solving which gives new epsilon} and $\bar{x}_{0|k}=x_k$, it follows that $\widetilde{\bf c}_k$ is a feasible solution to problem \eqref{eq:mpc_optimisation} at time $k$. 
On the other hand, if the chosen gain selection method gives $K_k\neq K_{k-1}$ and hence $\widetilde{P}_k \neq \widetilde{P}_{k-1}$, we can show that the guarantee of recursive feasibility is retained as follows. If gain selection method~\ref{scheme:less computation} is used, $\bar{\mu}_k$ defined in step (2) of Method \ref{scheme:less computation} satisfies the constraint in problem \eqref{eqn: maximisation in scheme 1}, which, together with equation \eqref{eq:feasibility retain}, implies a feasible solution to problem \eqref{eq:mpc_optimisation} exists at time $k$. If gain selection method~\ref{scheme:probably better} is used, $\bar{\mu}_k$ defined in step (3) of Method \ref{scheme:probably better} satisfies $\bar{\mu}_k \leq \tilde{\mu}$ and satisfies the constraint in problem \eqref{eqn:compute mu_tilde in scheme 2} due to monotonicity results in Lemmas \ref{lemma:monotonicity of Pbar and Phat} and \ref{lemma:min cost determined by terminal P}. This implies a feasible solution to problem \eqref{eq:mpc_optimisation} exists at time $k$.}

\subsection{Proof of Lemma~\ref{lemma:monotonicity of Pbar and Phat}}
\blue{In problem \eqref{eqn:DP cost:deterministic form},
the optimality of $x_0^\top \bigl((1-\mu)\bar{P}_{\infty}(\mu) + \mu\hat{P}_{\infty}(\mu) \bigr) x_0$ for all $x_0 \in \mathbb{R}^{n_x}$} implies
\begin{align*}
(1\!-\!\mu_1) \bar{P}_{\infty}(\mu_1) \!+\! \mu_1 \hat{P}_{\infty}(\mu_1) &\preccurlyeq (1 \!-\! \mu_1) \bar{P}_{\infty}(\mu_2) \!+\! \mu_1 \hat{P}_{\infty}(\mu_2), \\
(1\!-\!\mu_2) \bar{P}_{\infty}(\mu_2) \!+\! \mu_2 \hat{P}_{\infty}(\mu_2) &\preccurlyeq (1 \!-\! \mu_2) \bar{P}_{\infty}(\mu_1) \!+\! \mu_2 \hat{P}_{\infty}(\mu_1).
\end{align*}
Defining $\Delta \bar{P}$, $\Delta \hat{P}$ by $\Delta \bar{P} := \bar{P}_{\infty}(\mu_{2}) - \bar{P}_{\infty}(\mu_1)$, $\Delta \hat{P} := \hat{P}_{\infty}(\mu_{2}) - \hat{P}_{\infty}(\mu_1)$, we have
\begin{align}
-(1-\mu_1) \Delta \bar{P} &\preccurlyeq \mu_1 \Delta \hat{P}, \label{eq:DeltaPa} \\
(1-\mu_2) \Delta \bar{P} &\preccurlyeq -\mu_2 \Delta \hat{P}. \nonumber
\end{align}
Combining these inequalities yields
$\Delta \bar{P} \succcurlyeq0$ and $\Delta \hat{P} \preccurlyeq 0$.
\subsection{Proof of Lemma~\ref{lemma:concavity}}
Let $0 < \mu_1 \leq \mu_2 \leq \mu_3\leq 1$ with $\mu_3-\mu_2 = \mu_2 - \mu_1$, and let $\Delta^2\bar{P} = \bar{P}_{\infty}(\mu_3) - 2\bar{P}_{\infty}(\mu_2) + \bar{P}_{\infty}(\mu_1)$, $\Delta^2\hat{P} = \hat{P}_{\infty}(\mu_3) - 2\hat{P}_{\infty}(\mu_2) + \hat{P}_{\infty}(\mu_1)$ and $\Delta^2S = S(\mu_3) - 2 S(\mu_2) + S(\mu_1)$.
Then by definition we have
$
\Delta^2 S = (1-\mu_3) \Delta^2 \bar{P} + \mu_3 \Delta^2 \hat{P} -(\mu_3 - \mu_1) (\Delta \bar{P} - \Delta \hat{P}),
$
where $\Delta \bar{P} = \bar{P}_{\infty}(\mu_{2}) - \bar{P}_{\infty}(\mu_1)$, $\Delta \hat{P} = \hat{P}_{\infty}(\mu_{2}) - \hat{P}_{\infty}(\mu_1)$. But the optimality property of $\tr\bigl(Z S(\mu) \bigr)$ for all $Z=Z^\top\succcurlyeq 0$ implies
$(1-\mu_3) \bar{P}_{\infty}(\mu_3) + \mu_3 \hat{P}_{\infty}(\mu_3) \preccurlyeq (1 - \mu_3) \hat{P}_{\infty}(\mu_2) + \mu_3 \hat{P}_{\infty}(\mu_2)$, so
\[
(1-\mu_3) \bigl(\bar{P}_{\infty}(\mu_3)-\bar{P}_{\infty}(\mu_2)\bigr) \preccurlyeq -\mu_3 \bigl(\hat{P}_{\infty}(\mu_3) - \hat{P}_{\infty}(\mu_2) \bigr).
\]
Combining this inequality with (\ref{eq:DeltaPa}), we obtain
\[
(1-\mu_3) \Delta^2 \bar{P} + \mu_3 \Delta^2 \hat{P} - (\mu_3 - \mu_1) (\Delta \bar{P} - \Delta \hat{P}) \preccurlyeq 0,
\]
which implies $\Delta^2 S \preccurlyeq 0$ and hence $S(\mu)$ is midpoint concave for all $\mu\in(0,1]$.
Furthermore, $S(\mu)$ is positive definite (bounded below) for all $\mu\in(0,1]$. Therefore, $S(\cdot)$ is continuous and hence concave on $(0,1)$ in the sense that $\tr\bigl( Z S(\cdot) \bigr)$ is concave on $(0,1)$ for all $Z=Z^\top\succcurlyeq 0$ \cite[Section 72]{roberts1973convex}.
\subsection{Proof of Lemma~\ref{lemma:lipschitz continuity}}
\def\tV{\Sigma}
Let $\tV(\mu) := \gamma  (1-\mu) \bP_{\infty}(\mu) + \mu \hP_{\infty}(\mu)$, and $\Phi(\mu) := A+BL_{\infty}(\mu)$
and consider the effect of an infinitesimal change $\delta\mu$ in the value of $\mu\in(0,1)$, where $\delta\mu \ll  1-\gamma$ and $\delta\mu \ll 1-\mu$. Let $\Delta \Phi$, $\Delta \tV$, $\Delta L$, $\Delta \bP$, $\Delta \hP$ denote the corresponding changes in the values of $\Phi$, $\tV$, $L_{\infty}$, $\bar{P}_{\infty}$ and $\hat{P}_{\infty}$ respectively, so that
$\Delta \Phi = \Phi(\mu+\delta\mu) - \Phi(\mu)$,
$\Delta \tV = \tV(\mu+\delta\mu) - \tV(\mu)$,
$\Delta L = L_{\infty}(\mu+\delta\mu) - L_{\infty}(\mu)$,
$\Delta \bP = \bar{P}_{\infty}(\mu+\delta\mu) - \bar{P}_{\infty}(\mu)$, and
$\Delta \hP = \hat{P}_{\infty}(\mu+\delta\mu) - \hat{P}_{\infty}(\mu)$. To simplify notation we omit the argument $\mu$ for the remainder of the proof (so that $\Phi = \Phi(\mu)$, $\tV=\tV(\mu)$, $L_{\infty}=L_{\infty}(\mu)$, etc.).
Lemma~\ref{lemma:concavity} implies that $(1-\mu)\bar{P}_{\infty} + \mu \hat{P}_{\infty}$ is Lipschitz continuous on $(0,1)$, so $(1-\mu) \Delta\bP + \mu \Delta\hP - \delta\mu (\bar{P}_{\infty}(\mu+\delta \mu) - \hat{P}_{\infty}(\mu+\delta \mu)) = O(\delta \mu)$ and
\begin{multline}
\hspace{-4mm} O(\delta\mu) = (1-\mu) \Delta\bP + \mu \Delta\hP
= (\Phi + \Delta\Phi)^\top ( \tV + \Delta \tV) (\Phi + \Delta\Phi) \\
- \Phi^\top \tV \Phi
+ \mu (L_{\infty} + \Delta L)^\top R (L_{\infty}+ \Delta L) - \mu L_{\infty}^\top R L_{\infty}.\label{eq:drhs}
\end{multline}
But
$L_{\infty} + \Delta L = -\bigl[(\mu + \delta\mu) R + B^\top (\tV + \Delta \tV) B\bigr]^{-1}B^\top (\tV + \Delta \tV)A$
implies
\begin{multline}
L_{\infty}+\Delta L = L_{\infty}-\Gamma^{-1}( B^\top \Delta \tV \Phi + R L_{\infty} \delta\mu) \\
= L_{\infty} - \Gamma^{-1} B^\top \Delta \tV \Phi + O(\delta\mu), \label{eq:dL}
\end{multline}
where $\Gamma = (\mu + \delta\mu) R + B^\top (\tV + \Delta \tV) B$, and
\begin{equation}
\Phi + \Delta \Phi = \Phi - B\Gamma^{-1}B^\top \Delta \tV \Phi + O(\delta\mu).
\label{eq:dPhi}
\end{equation}
Combining (\ref{eq:dL}) and (\ref{eq:dPhi}) with (\ref{eq:drhs}) and using $L_{\infty} = -(\mu R + B^\top \tV B)^{-1} B^\top \tV A$, we obtain
\begin{align*}
O(\delta\mu) &= \Phi^\top \Delta \tV \Phi \\
&+ \Phi^\top \Delta \tV B \Gamma^{-1} (B^\top \tV B + \mu R - 2\Gamma )\Gamma^{-1} B^\top \Delta \tV \Phi
\\
&+ \Phi^\top \Delta \tV B \Gamma^{-1} B^\top \Delta \tV B \Gamma^{-1} B^\top \Delta \tV \Phi
\\
&= \Phi^\top \Delta \tV \Phi - \Phi^\top \Delta \tV B \Gamma^{-1} B^\top \Delta \tV \Phi
\\
&- \Phi^\top \Delta \tV B \Gamma^{-1} R \Gamma^{-1} B^\top \Delta \tV \Phi \delta\mu ,
\end{align*}
and hence
\[
\Phi^\top \Delta \tV \Phi - \Phi^\top \Delta \tV B \Gamma^{-1} B^\top \Delta \tV \Phi
= O(\delta\mu) .
\]
But
$\tV = (1-\mu)\bar{P}_{\infty} + \mu \hat{P}_{\infty} - (1-\gamma) (1-\mu) \bar{P}_{\infty}$ implies
$\Delta \tV = O(\delta\mu) -(1-\gamma) (1-\mu) \Delta\bP$, so
\[
(1\!-\gamma) (1\!-\!\mu) \Phi^\top \! \bigl[ \Delta\bP \!+\! (1\!-\gamma) (1\!-\!\mu)\Delta\bP B \Gamma^{-1} \! B^\top \! \Delta \bP\bigr]\! \Phi
\!=\! O(\delta\mu) .
\]
This implies $\Phi^\top \Delta\bP\Phi = O(\delta\mu)$ since $\delta\mu \ll 1-\gamma$, $\delta\mu \ll 1-\mu$ and $\Gamma\succcurlyeq0$, and since $\Delta \bP \succcurlyeq0$ by Lemma~\ref{lemma:monotonicity of Pbar and Phat}. Furthermore, $\Delta \bP$ is symmetric so we must have $\Delta \bP \Phi = O(\delta\mu)$ and hence $\Delta \tV \Phi = O(\delta\mu)$. From (\ref{eq:dL})-(\ref{eq:dPhi}) it follows that $\Delta L = O(\delta\mu)$ and $\Delta \Phi = O(\delta\mu)$, and hence the solutions, $\bP_{\infty}(\cdot)$, $\hP_{\infty}(\cdot)$ of the Lyapunov equations \eqref{eq:steady state Pbar}-\eqref{eq:steady state Phat} are Lipschitz continuous on $(0,1)$.
\subsection{Proof of Lemma \ref{lemma:prob_mubound}}
We prove this lemma by contradiction.
Suppose that a pair of $\delta > 0$ and $p_\delta > 0$ does not exist such that \eqref{eq:prob_mubound} holds. In this case, for some
$z\in\mathbb{R}^{Nn_u}$ and $\mu\in[\bar{\mu}_0,1]$ we must have
\begin{equation}\label{eq:prob_mubound_contra}
\prob{
\bigl\|z + \bigl(W_{cc}(\mu)\bigr)^\dagger W_{cx}(\mu) \omega_{k} \|^2_{W_{cc}(\mu)}
\geq \delta } < p_{\delta}
\end{equation}
for all $\delta > 0$ and all $p_{\delta} > 0$, which implies that $W_{cx}(\mu) \omega_{k} + W_{cc}(\mu) z = 0$ with probability~1.
Here $W_{cx}(\mu)$ must be non-zero since otherwise $\mathbf{c}^o(x_k)=0$, implying that by its definition in \eqref{eqn:min in scheme 1 to make space for K} the feedback gain used is optimal with respect to $\sum_{i=0}^{\infty}\gamma^i \|C\bar{x}_i\|^2_Q$
and then $L_\infty(\mu) = L_{\infty}(0)$ would be obtained, which contradicts the fact that $\mu\geq\bar{\mu}_0\geq\mu_1$ and the
statement that $L_{\infty}(\mu) \neq L_{\infty}(0)$ $\forall \mu \geq \mu_1$
in Section \ref{section:multiobjective optimisation and dynamic programming}.
Furthermore, $z$ and $\mu$ are by assumption independent of the realisation $\omega_{k}$. Therefore (\ref{eq:prob_mubound_contra}) contradicts the assumption in \eqref{eq:disturbance statistics} that $\EE \{\omega_{k}\omega_{k}^\top\}$ is positive definite implying $\omega_k$ is not a constant vector or in any subspaces of $\mathbb{R}^{n_x}$, and hence (\ref{eq:prob_mubound}) must hold for some $\delta > 0$ and $p_\delta >0$.
\subsection{Proof of Theorem \ref{theorem:almost sure converge}}
We first show convergence of $\bar{\mu}_k$ to 1 in probability (i.e.\
$\lim_{k\to\infty}\PP\{\bar{\mu}_k = 1\}=1$) by splitting an infinite horizon into intervals of $N_f$ time steps and providing an upper bound on $\PP \{\bar{\mu}_k< 1\}$, which is parameterised by $p_\delta$, $N_f$ and $k$, and which converges to 0 as $k\to\infty$.
Let $N_f$ be defined in terms of a $\delta>0$ satisfying (\ref{eq:prob_mubound}) for some $p_\delta>0$ by
\[
N_f = \biggl\lceil \frac{\gamma}{1-\gamma} \frac{ \tr \left(\Omega \bP_{\infty}(1)-\Omega \bP_{\infty}(\bar{\mu}_0)\right)}{\delta} \biggr\rceil
\]
where it is assumed that $\bar{\mu}_0<1$,
and let $\mathcal{E}_k$ denote the event that
\[
\|z_{k-1} + \bigl(W_{cc}(\bar{\mu}_{k-1})\bigr)^\dagger W_{cx}(\bar{\mu}_{k-1}) \omega_{k-1} \|^2_{W_{cc}(\bar{\mu}_{k-1})} \geq \delta.
\]
Under Assumption \ref{assumption:can solve continuous optimisation}, Lemma \ref{lemma:lipschitz continuity} implies that before $\bar{\mu}_k$ reaches $1$, \eqref{eq:the key constraint in scheme 1} is satisfied with equality and every possible increment in $\bar{\mu}_k$ is attained if \eqref{eq:RHS(w_(k-1))} is positive. It follows that $\bar{\mu}_{N_f} = 1$
if $\mathcal{E}_k$ occurs for $k=1,\ldots,N_f$.
Furthermore,
\begin{align*}
&\prob{ \mathcal{E}_1 \cap \cdots \cap \mathcal{E}_{N_f} } \\
&\hspace{-1.0mm}=
\prob{ \mathcal{E}_1 }
\prob{ \mathcal{E}_2 | \mathcal{E}_1 }
\cdots
\prob{ \mathcal{E}_{N_f} | \mathcal{E}_1 \cap \mathcal{E}_2 \cap \cdots \cap \mathcal{E}_{N_f-1}},
\end{align*}
where $\prob{ \mathcal{E}_1 } \geq p_\delta$ and
\begin{align*}
&\prob { \mathcal{E}_k | \mathcal{E}_1 \cap \cdots \cap \mathcal{E}_{k-1}} \\
&\hspace{-1.0mm}\geq \hspace{-1.0mm}
\inf_{z\in\mathbb{R}^{N\!n_u}\!,   \mu\in [\bar{\mu}_0,1]} \!
\mathbb{P} \big\{
\bigl\|z \!+\! \bigl(W_{cc}(\mu)\bigr)^\dagger W_{cx}(\mu) \omega_{k-1} \bigr\|^2_{W_{cc}(\mu)} \!\geq\! \delta \big\} \\
&\hspace{-1.0mm}\geq
p_\delta
\end{align*}
for $k=2,\ldots, N_f$. Also, these conditional probabilities are well defined by Lemma \ref{lemma:prob_mubound}. Therefore,
\[
\prob{ \bar{\mu}_{N_f} < 1 } \leq  1 - p_\delta^{N_f},
\]
and more specifically we rewrite it as
\[
\prob{ \bar{\mu}_{N_f} < 1 \big| \bar{\mu}_0 <1 } \leq  1 - p_\delta^{N_f}.
\]
Similarly, given $\bar{\mu}_{(i-1)N_f}<1$ for $i=2, \ldots$, we have
\begin{align*}
&\prob{\bar{\mu}_{iN_f}=1} \geq \!
\Big( \inf_{z\in\mathbb{R}^{N\!n_u}\!, \  \mu\in [\bar{\mu}_0,1]}\!
\mathbb{P} \{ \star \geq \delta_i \} \Big)^{N_f} \\
&\geq \!
\Big( \inf_{z\in\mathbb{R}^{N\!n_u}\!, \ \mu\in [\bar{\mu}_0,1]}\!
\mathbb{P} \{ \star \geq \delta \} \Big)^{N_f} \geq
p_\delta^{N_f},
\end{align*}%
where $\star$ here denotes $\bigl\|z \!+\! \bigl(W_{cc}(\mu)\bigr)^\dagger W_{cx}(\mu) \omega_{iN_f-1} \bigr\|^2_{W_{cc}(\mu)}$, and $\delta_i$ is the counterpart of $\delta$ for $\bar{\mu}_{iN_f}$ that is sufficient for $\bar{\mu}_k$ to reach $1$ within $N_f$ time steps from time $k=(i-1)N_f$ and is smaller than $\delta$ since the sequence $\{\bar{\mu}_k\}_{k=0}^{\infty}$ is monotonically non-decreasing. Then
\[
\prob{\bar{\mu}_{iN_f}<1 \big| \bar{\mu}_{(i-1)N_f}<1} \leq 1-p_\delta^{N_f},\quad \forall~ i=2, \ldots .
\]
For given $k>N_f$, we choose integers $j_1,j_2,\ldots,j_{\lfloor k/N_f \rfloor}$ so that $j_1 = N_f$, $j_{\lfloor k/N_f \rfloor} \leq k$ and $j_{i+1} - j_i = N_f$ for all $i$. Then $\bar\mu_k < 1$ only if $\bar\mu_{j_i} < 1$ for all $i=1,\ldots,\lfloor k/N_f \rfloor$ and hence
\begin{align}
&\prob{\bar{\mu}_k<1} \nonumber \\
&\hspace{-1mm}\leq \hspace{-1mm}
\mathbb{P}\{\bar{\mu}_{j_1}\!\!<\!1\} \mathbb{P}\{ \bar{\mu}_{j_2}\!\!<\!1 \big| \bar{\mu}_{j_1}\!\!<\!1\} \!\cdots\! \mathbb{P}\{\bar{\mu}_{j_{\lfloor\! k\!/\!N_f \!\rfloor}}\!\!<\!1 \big| \bar{\mu}_{j_{\lfloor\! k\!/\!N_f \!\rfloor-1}}\!\!<\!1 \} \nonumber \\
&\hspace{-1mm}\leq \hspace{-1mm}
\left(1-p_\delta^{N_f}\right)^{\lfloor k/N_f \rfloor}. \label{eq:prob Fk}
\end{align}
It follows that $\mathbb{P} \left\{ \bar\mu_k = 1 \right\} \geq 1-(1- p_\delta^{N_f})^{\lfloor k/N_f \rfloor}$, which implies $\lim_{k\to\infty} \mathbb{P} \left\{ \bar\mu_k =1 \right\} = 1$.
To complete the proof,
we use the Borel-Cantelli lemma to show the almost sure convergence of $\bar{\mu}_k$ to 1, that is, $\prob{ \lim_{k\to\infty} \bar\mu_k = 1 } = 1$.
Let $\mathcal{F}_k$ denote the event that $\bar\mu_k < 1$, then (\ref{eq:prob Fk}) ensures that
 \[
\sum_{k=1}^\infty \prob{\mathcal{F}_k} \leq \sum_{k=1}^\infty ( 1- p_\delta^{N_f})^{\lfloor k/N_f \rfloor} = N_f p_\delta^{-N_f} -1 < \infty.
\]
Therefore the Borel-Cantelli lemma implies that $\prob{ \bigcap_{k=1}^\infty \bigcup_{j=k}^\infty \mathcal{F}_j } = 0$. But $\bar\mu_k < 1$ only if $\bar\mu_{k-1} < 1$, so $\mathcal{F}_k \subseteq \mathcal{F}_{k-1}$ and $ \bigcup_{j=k}^\infty \mathcal{F}_j = \mathcal{F}_k$. Therefore, $\prob{ \lim_{k\to\infty}  \mathcal{F}_k } = 0$ or equivalently $\prob{ \lim_{k\to\infty} \bar\mu_k = 1 } = 1$.
\subsection{Proof of Theorem \ref{theorem:stability result}}
Given initial feasibility at time $k=0$, step (ii) of Algorithm \ref{algorithm:SPMC algorithm} ensures that problem \eqref{eq:mpc_optimisation} is always feasible for time $k=1,2,\ldots$.
  From Lemma \ref{theorem:theorem for recursive feasibility}, the vector $\widetilde{\mathbf{c}}_{k+1}$ provides a feasible but possibly suboptimal solution at time $k+1$. Hence by optimality we necessarily have
  \[
    J^\ast(x_{k+1}, K_{k+1}) \leq
    J\bigl(x_{k+1}, K_k, \widetilde{\mathbf{c}}_{k+1}\bigr),
  \]
and since this inequality holds for every realisation of $\omega_{k}$, by taking expectations conditioned on the state $x_k$ we obtain
\begin{equation}
 \mathbb{E}_k \{  J^\ast(x_{k+1}, K_{k+1}) \} \!\leq\!
    \mathbb{E}_k \{ J\bigl(x_{k+1}, K_k, \widetilde{\mathbf{c}}_{k+1}\bigr)\}.
\label{eqn:expected_cost_diff}
\end{equation}
From \eqref{eqn:feasible state sequence} we have the feasible sequence
\[
  \bar{x}_{i|k+1} = \bar{x}^\ast_{i+1|k} + \Phi_k^i \omega_k , \qquad i = 0,\ldots,N ,
\]
and from \eqref{eqn:lyap P_k} and \eqref{eqn:expected_cost_diff} it follows that
\begin{multline}
  \Ex[k]{ J^\ast (x_{k+1}, K_{k+1})} \leq J^\ast (x_k, K_k) \\
   - \norm{x_k}_Q^2 - \norm{ u_k }_R^2 + \tr\bigl(\Omega P_k\bigr) .
  \label{eqn:cost comparison in mean, one optimal and one feasible at time k+1}
\end{multline}

Summing both sides of this inequality over $k\geq 0$ after taking expectations given information available at time ${k=0}$, and making use of the property that $\Ex[0]{\Ex[k]{J^\ast (x_{k+1}, K_{k+1})}} = \Ex[0]{J^\ast (x_{k+1}, K_{k+1})}$, give the first inequality of \eqref{eqn:quadratic stability result}. Note that $K_k$ and $P_k$ depend on $x_k$, which implies that they are random variables, but they are uncorrelated with $\omega_k$. Moreover, the second inequality in \eqref{eqn:quadratic stability result} follows from the choice of $K_k$ in step (ii) of Algorithm~\ref{algorithm:SPMC algorithm} and Lemma~\ref{lemma:monotonicity of Pbar and Phat}.
\subsection{Proof of Lemma \ref{thm:epsilon}}
Since \eqref{eqn: updating epsilon: optimisation by solving which gives new epsilon} is equivalent to
\eqref{eqn:epsilon of next time instant:expression of new epsilon using results obtained at current time}, expanding the terms in \eqref{eqn:epsilon of next time instant:expression of new epsilon using results obtained at current time} yields
\begin{align}
& \varepsilon_{k+1} \nonumber \\
&\!=
 \sum_{i=0}^{N-1} \gamma^i \bigl\| C \bar{x}_{i+1|k}^* \bigr\|^2 + \gamma^N \bigl\| \bar{x}^*_{N+1|k}\bigr\|^2_{\tP_k}  + \frac{\gamma}{1-\gamma} \tr\bigl( \Omega \tP_k \bigr)  \nonumber \\
& \! + \! \sum_{i=0}^{N-1} \gamma^i \bigl\| C \Phi^i_k \omega_k
\bigr\|^2 + \gamma^N \bigl\| \Phi_k^N \omega_k\bigr\|^2_{\tP_k}    \nonumber \\
& \!  + \sum_{i=0}^{N-1}  2 \gamma^i \left( \Phi^i_k \omega_k \right)^\top \! C^\top C \bar{x}_{i+1|k}^*
 \!+\!  2 \gamma^N \left( \Phi^N_k \omega_k \right)^\top \! \tP_k \, \bar{x}_{N+1|k}^* ,\label{eqn:full expression e(k+1)}
\end{align}
where $\bar{x}^\ast_{i|k}$ is given by~\eqref{eqn:open loop system: optimal state trajectory, first mode}-\eqref{eqn:open loop system: optimal state trajectory, second mode} and $\omega_k$ is the realisation of the disturbance at time $k$.
From \eqref{eq:Plyap} and $\bar{x}^*_{N+1|k} = \Phi_k\bar{x}^*_{N|k}$, the sum of the first three terms on the RHS of \eqref{eqn:full expression e(k+1)} is
\begin{multline*}
\gamma^{-1} \Big( \sum_{i=0}^{N-1} \gamma^i \| C \bar{x}_{i|k}^\ast \|^2  + \gamma^N \| \bar{x}_{N|k}^\ast \|^2_{\tP_k} + \\
\frac{\gamma}{1-\gamma} \tr \bigl(\Omega \tP_k\bigr)  - \|C x_k\|^2 \Big)
- \tr \bigl(\Omega \tP_k\bigr),
\end{multline*}
and from \eqref{eq:Plyap} the sum of the next two terms is $\omega_k^\top  \tP_k \omega_k$.
Noting that $K_k$, $\tP_k$ and $x_k$ are independent of $\omega_k$, taking the expectation of $\varepsilon_{k+1}$ conditioned on information available at time $k$ therefore gives
\begin{align*}
&\gamma\Ex[k]{\varepsilon_{k+1}}=\\
&  \sum_{i=0}^{N-1} \gamma^i \|C \bar{x}_{i|k}^\ast \|^2  \!+\! \gamma^N \| \bar{x}_{N|k}^\ast \|^2_{\tP_k}    \!+\! \frac{\gamma}{1-\gamma} \tr \bigl(\Omega \tP_k\bigr)  \!-\! \|C x_k\|^2              .
\end{align*}
This equation, together with feasibility of the sequence $\{\bar{x}^\ast_{i|k}\}_{i=0}^{N}$ at time $k$, proves \eqref{eq:epsilon}.

\bibliographystyle{IEEEtran}
\bibliography{IEEEabrv,reference}

\end{document}